\documentclass[a4paper,11pt]{article}
\pdfoutput=1 

\usepackage{jheppub} 

\usepackage[T1]{fontenc} 
\usepackage{subcaption}
\graphicspath{ {graphs/} }

\preprint{PUPT-2587,\\
	\phantom{~} \hfill YITP-SB-2019-17}

\title{\boldmath $\mathbf{AdS_3\times S^3}$ Tree-Level Correlators: \\ Hidden Six-Dimensional Conformal Symmetry}

\author[]{Leonardo Rastelli$^{D1, K3}$,}
\author[]{Konstantinos Roumpedakis$^{D1}$,}
\affiliation[]{$^{D1}$YITP, 
Stony Brook University, Stony Brook, NY 11794, USA}
\author[]{Xinan Zhou$^{D5}$}
\affiliation[]{$^{D5}$Princeton Center for Theoretical Science, Princeton University, Princeton, NJ 08544, USA}
\affiliation[]{$^{K3}$CERN, Theoretical Physics Department, 1211 Geneva 23, Switzerland}

\abstract{We  revisit the calculation of holographic correlators in $AdS_3$. We  develop new methods to evaluate  exchange Witten diagrams, resolving some technical difficulties that prevent a straightforward application of  the methods used in higher dimensions. We perform detailed calculations in the  $AdS_3 \times S^3 \times K3$ background. 
  We find strong evidence that   four-point tree-level correlators of KK modes of the tensor multiplets  enjoy a hidden 6d conformal symmetry. The correlators can all  be packaged into a single generating function, related to the 6d flat space superamplitude. This generalizes an analogous structure found in $AdS_5 \times S^5$ supergravity.}

\begin{document}
\maketitle
\flushbottom
\section{Introduction}

Correlation functions of local operators are basic observables in holographic CFTs, and as such have been intensely studied  since  the early days of AdS/CFT. Only recently however have truly efficient computational methods
been developed. Broadly speaking, these new techniques are inspired by the modern ``on-shell'' approach to perturbative gauge theory amplitudes. One focusses on the full holographic correlator, which is a much simpler and more rigid object than individual Witten diagrams. Correlators can be strongly constrained and sometimes completely determined by imposing symmetries and other consistency requirements. 

In the paradigmatic AdS/CFT example, namely the dual pair of  $\mathcal{N}=4$ super Yang-Mills theory  and IIB string theory on $AdS_5\times S^5$, this new approach has led to a compelling conjecture for {\it all} one-half BPS four-point correlators in the tree-level supergravity limit \cite{Rastelli:2016nze,Rastelli:2017udc}.\footnote{Some highly non-trivial checks of this conjecture have been performed in \cite{Arutyunov:2017dti,Arutyunov:2018tvn,Arutyunov:2018neq}  by explicit supergravity calculations.} From these tree-level  correlation functions, one can extract the a wealth of  $O(1/N^2)$ CFT data about non-protected double-trace operators \cite{Alday:2017xua,Aprile:2017bgs,Aprile:2017xsp,Aprile:2018efk,Caron-Huot:2018kta}.    In turn, these  data serve as input  in the ``AdS unitarity method'' \cite{Aharony:2016dwx} to yield one-loop results and thus  $O(1/N^4)$ anomalous dimensions \cite{Alday:2017xua,Aprile:2017bgs,Aprile:2017qoy,Alday:2017vkk}.\footnote{One can further consider stringy corrections to four-point functions, see, {\it e.g.}, \cite{Goncalves:2014ffa,Alday:2018pdi,Alday:2018kkw,Binder:2019jwn} for work in this direction.}
 The conjectured tree-level correlators \cite{Rastelli:2016nze,Rastelli:2017udc}, and the CFT data that can be extracted from them,  take a remarkably simple form, which has been interpreted \cite{Caron-Huot:2018kta} as arising from a ``hidden''  (and rather mysterious) 10d conformal symmetry.

 This success story has been replicated, at least partially, for other supergravity backgrounds.  The techniques developed for $AdS_5 \times S^5$  have been generalized to $AdS_7$ \cite{Rastelli:2017ymc,Zhou:2017zaw,Heslop:2017sco,Zhou:2018ofp,Chester:2018dga,Abl:2019jhh}, $AdS_6$ \cite{Zhou:2018ofp} and $AdS_4$ \cite{Zhou:2017zaw,Chester:2018lbz,Chester:2018aca,Binder:2018yvd} backgrounds with maximal or half-maximal supersymmetry, leading to many interesting new results.  By contrast,  $AdS_3$ backgrounds are more challenging and have so far defied our efforts. The purpose of the present work is to  remedy this situation.
  
  The greater technical difficulty of the $AdS_3$ case can be traced to the chiral nature of boundary correlators (and, dually, of bulk Witten diagrams).
  Four-point correlators in a 2d CFT are  functions\footnote{We have in mind the usual kinematic setup of choosing a conformal frame where the positions of three operators are fixed, so that the correlator depends only on the coordinates of forth operator.} of the usual holomorphic coordinates on the plane, $z$ and $\bar z$, but unlike the situation in higher dimensional CFTs, there is no requirement of symmetry under the exchange of $z$ and $\bar z$.   This chiral character limits the applicability of the  Mellin  formalism \cite{Mack:2009mi,Penedones:2010ue,Paulos:2011ie,Fitzpatrick:2011ia}, so the Mellin bootstrap approach of \cite{Rastelli:2016nze,Rastelli:2017udc,Zhou:2017zaw,Rastelli:2017ymc,Zhou:2018ofp}  does not generalize immediately to $AdS_3$. On the other hand, the ``position space''
  method developed in \cite{Rastelli:2016nze,Rastelli:2017udc} can be applied, but requires some extra work. In the position space approach, one writes an ansatz for the holographic correlator as a sum of Witten
  diagrams, taking into account only general selection rules that follows from the structure of the supergravity theory, but with arbitrary coefficients. The coefficients are then fixed by imposing superconformal Ward identities. In the $AdS_3$ case, there are new types exchange Witten diagrams for Chern-Simons bulk interactions, and the standard techniques of \cite{DHoker:1999mqo} need to be suitably generalized. There are further subtle issues in the massless limit of the bulk exchanged field, even for the standard Maxwell and gravity cases. We resolve all of these technical difficulties in the present paper.

While our new results for exchange Witten diagrams in $AdS_3$ are of general applicability, we will focus on performing detailed calculations in the best studied background, namely
$AdS_3\times S^3\times K3$, which arises in the near-horizon limit of the D1-D5 system.\footnote{The generalization to $AdS_3\times S^3\times T^4$ is straightforward.} 
By reducing IIB supergravity on $K3$, whose size is taken to be much smaller than the (common) radius of curvature of $AdS_3$ and $S^3$, one obtains 6d (2,0) supergravity coupled to 21 tensor multiplets.
Upon further reduction  on $S^3$, the gravity multiplet and the tensor multiplets  give rise to infinite Kaluza-Klein towers of  one-half BPS multiplets of the $PSU(1, 1 |2) \times PSU(1, 1, |2)$ superalgebra.
Prior to our work, some partial results had been obtained using indirect methods \cite{Galliani:2017jlg,Bombini:2017sge,Giusto:2018ovt,Bombini:2019vnc}. For example, the four-point function of the KK modes with  lowest conformal dimension was conjectured from a limit of the light-light-heavy-heavy correlators \cite{Giusto:2018ovt}.

 There is also a clear physical incentive to revisit  $AdS_3 \times S^3$ holographic correlators, beyond the mere demonstration that our previous methods can be generalized to this more difficult case. 
 Do these correlators exhibit a hidden conformal symmetry analogous to the one found for $AdS_5 \times S^5$ case  \cite{Caron-Huot:2018kta}? While a conceptually satisfactory explanation is still lacking, in the  10d case such a symmetry appears to hinge crucially on  a few facts. First, the $AdS_5 \times S^5$  metric is conformally flat, a feature shared by the $AdS_3 \times S^3$ background but not (for example) by the maximally supersymmetric M-theory cases, namely\footnote{Indeed, in the M-theory cases, the radii of the AdS and sphere factors are different. In the case of $AdS_7\times S^4$ one can immediately see that a putative 11d conformal symmetry would be structurally incompatible with the explicit results of \cite{Rastelli:2017ymc}. 
 } $AdS_7 \times S^4$ and $AdS_4 \times S^7$. Second, the flat space 10d superamplitude in IIB supergravity contains a kinematic factor $G_N^{(10)}  \delta^{16}(Q)$, which can in some heuristic sense be regarded
 as a  dimensionless coupling. An analogous power-counting applies to superamplitudes in 6d $(2, 0)$ supergravity, where the kinematic factor $G_N^{(6)}  \delta^{8}(Q)$ is again dimensionless. 
Third,  the  four-point  superamplitude  in 10d flat space IIB supergravity enjoys an accidental 10d conformal symmetry. The 10d amplitude can be viewed as a generating function of all
four-point correlators of the full tower of KK modes on $AdS_5 \times S^5$  \cite{Caron-Huot:2018kta}. The situation in 6d $(2, 0)$ flat space supergravity is more elaborate. As we have mentioned,  there are {\it two} relevant supermultiplets, 
the graviton and tensor multiplets. As it turns out, the superamplitude with four external tensors enjoys an accidental 6d conformal symmetry!\footnote{We were not able to find in the literature fully explicit expressions for amplitudes
involving external supergravitons in 6d  $(2, 0)$ supergravity (see \cite{Heydeman:2018dje} for the state of the art), but we suspect that they do not enjoy such a symmetry.}

By analogy with the $AdS_5 \times S^5$ case, it seems natural to anticipate that 
that all four-point correlators of  tensor multiplet KK modes in $AdS_3 \times S^3$ can be packaged into a single 6d object. We find strong evidence that this is indeed the case. Our strategy is to develop a systematic 
position space method similar to the one used in \cite{Rastelli:2016nze,Rastelli:2017udc,Rastelli:2017ymc}. New ingredients in $AdS_3$ include a derivation of the superconformal Ward identity, and the computation  of $AdS_3$ exchange Witten diagrams that require a generalization of the existent  techniques.
Using this method we compute equal-weight four-point functions of one-half BPS operators that arise as KK modes of the 6d tensor multiplets. We have obtained results for operators with conformal dimensions $\Delta =1, 2, 3, 4$.  Our result for $\Delta = 1$ reproduces the recent conjecture of \cite{Giusto:2018ovt}.
We also discuss an independent method in Mellin space. The Mellin space method for $AdS_3\times S^3\times K3$ is not as powerful as in $AdS_5\times S^5$ and $AdS_7\times S^4$, but is still very useful to illustrate the analogy between holographic correlators and scattering amplitudes, which plays a crucial role in formulating a guess for the master formula. 

From these concrete examples of correlators, we observe nontrivial evidence of a six-dimensional hidden conformal symmetry. Assuming that such a symmetry persists for arbitrary external weights, it is immediate to write down a simple generating function of all four-point correlators of tensor KK modes
 by replacing the $x_{ij}^2$ in the lowest-weight four-point function with six dimensional distances.   On the other hand, 
such a hidden conformal symmetry is {\it not} present in the  four-point functions of scalar one-half BPS operators that arise from the 6d gravity multiplet, as we have checked using the position space method.
We hope that the new data obtained here will stimulate a better understanding of the nature of the hidden conformal symmetry, in both the $AdS_5 \times S^5$ and the $AdS_3 \times S^3$ cases.

\medskip

The rest of the paper is organized as follows. In Section \ref{Secscfkinematics} we discuss the superconformal kinematics of scalar one-half BPS four-point functions. In Section \ref{Secpositionspace} we introduce the position space method for $AdS_3$ and compute several examples of four-point functions. In Section \ref{SecMellinspace} we provide a different perspective in Mellin space. In Section \ref{Sechiddensymm}, we point out the existence of a six-dimensional hidden conformal symmetry. Using this symmetry we conjecture a formula for all one-half BPS four-point functions. We conclude in Section \ref{Secfuturedirec} by mentioning some future directions. The paper also includes three appendices to which we have relegated various technical details.

\medskip
{\bf Note:} As we were about to submit this paper to the arXiv, we learnt of an upcoming work \cite{Giusto} that obtains $AdS_3 \times S^3$ four-point tree-level correlators  with pairwise equal weights by generalizing the approach of \cite{Giusto:2018ovt}.

\section{Superconformal kinematics}\label{Secscfkinematics}

Let us start with the constraints of superconformal invariance. We focus on the one-half BPS local operators $\mathcal{O}_k^{\alpha_1\ldots\alpha_k,\dot{\alpha}_1\ldots\dot{\alpha}_k}(x)$ with $\alpha_i, \dot{\alpha}_i=1,2$, in the $(j,\bar{j})=\left(\frac{k}{2},\frac{k}{2}\right)$ representation of $SU(2)_L\times SU(2)_R$. These operators have protected conformal dimensions $(h,\bar{h})=\left(\frac{k}{2},\frac{k}{2}\right)$. The Kaluza-Klein reduction on $AdS_3\times S^3$ of 6d $(2,0)$ supergravity coupled to 21 tensor multiplets leads to two different types of one-half BPS scalar operators. The first kind originates from the anti-self-dual tensors with $k=1,2,\ldots$, and they transform in the vector representation of the $SO(21)$ flavor symmetry. The second kind comes from 6d supergravity fields with $k=2,3,\ldots$, and they are neutral under the flavor symmetry. In this work, we focus on correlation function of half-BPS operators in the tensor multiplet, although the superconformal constraints are the same for both types of operators. 

To begin with, it is convenient to keep track of the R-symmetry structure by contracting all the indices with auxiliary spinors $v_{\alpha}$, $\bar{v}_{\dot{\alpha}}$
\begin{equation}
\mathcal{O}_k^I(x;y,\bar{y})=\mathcal{O}_k^{I,\alpha_1\ldots\alpha_k,\dot{\alpha}_1\ldots\dot{\alpha}_k}v_{\alpha_1}\ldots v_{\alpha_k}\bar{v}_{\dot{\alpha}_1}\ldots \bar{v}_{\dot{\alpha}_k}\;.
\end{equation}
 We have exploited the fact that the spinors are automatically ``null''
\begin{equation}
\epsilon^{\alpha \beta}v_\alpha v_\beta=\epsilon^{\dot{\alpha}\dot{\beta}}\bar{v}_{\dot{\alpha}}\bar{v}_{\dot{\beta}}=0\;,
\end{equation}
and the one-half BPS operator is symmetric and traceless (with respect to the $\epsilon$ tensor) in $\alpha_i$ and $\dot{\alpha}_j$. We note that rescaling preserves the null property of the spinors. This allows us to parameterize the spinors as
\begin{equation}
v=\left(\begin{array}{c}1 \\ y \end{array}\right)\;,\quad \bar{v}=\left(\begin{array}{c}1 \\ \bar{y} \end{array}\right)\;.
\end{equation}
The goal of this paper is to calculate the four-point function 
\begin{equation}
G_{k_1k_2k_3k_4}^{I_1I_2I_3I_4}=\langle\mathcal{O}_{k_1}^{I_1}\mathcal{O}_{k_2}^{I_2}\mathcal{O}_{k_3}^{I_3}\mathcal{O}_{k_4}^{I_4}\rangle~.
\end{equation}
This is then a function of both the spacetime as well as the R-symmetry coordinates. Covariance under the conformal and R-symmetry implies that it is really a function of the cross ratios
\begin{equation}
z=\frac{z_{12}z_{34}}{z_{13}z_{24}}\;,\;\; \bar{z}=\frac{\bar{z}_{12}\bar{z}_{34}}{\bar{z}_{13}\bar{z}_{24}}\;, \;\;\alpha=\frac{y_{13}y_{24}}{y_{12}y_{34}}\;,\;\;\bar{\alpha}=\frac{\bar{y}_{13}\bar{y}_{24}}{\bar{y}_{12}\bar{y}_{34}}\;,
\end{equation}
and $z_{ij}\equiv z_i-z_j$, $y_{ij}\equiv y_i-y_j$, {\it etc}. Hence we can write it as
\begin{equation}
G^{I_1I_2I_3I_4}_{k_1k_2k_3k_4}=\mathbf{K}\, \mathcal{G}^{I_1I_2I_3I_4}_{k_1k_2k_3k_4}(z,\bar{z};\alpha,\bar{\alpha})~,
\end{equation} 
where the kinematic factor $\mathbf{K}$ is given by 
\begin{equation}
\mathbf{K}=\prod_{i<j}\left(\frac{y_{ij}\bar{y}_{ij}}{z_{ij}\bar{z}_{ij}}\right)^{\gamma^0_{ij}}\left(\frac{y_{12}y_{34}\bar{y}_{12}\bar{y}_{34}}{z_{12}z_{34}\bar{z}_{12}\bar{z}_{34}}\right)^L\;.
\end{equation}
Following the convention in \cite{Rastelli:2016nze,Rastelli:2017udc}, without loss of generality, we order the weights as $k_1\geq k_2\geq k_3\geq k_4$. There are two cases: I) $k_1+k_4\leq k_2+k_3$, II) $k_1+k_4>k_2+k_3$ where 
\begin{equation}
L=k_4\;\;\text{(case I)}\;,\quad\quad\quad L=\frac{k_2+k_3+k_4-k_1}{2}\;\;\text{(case II)}\;.
\end{equation}
The various $\gamma_{ij}^0$ are given by
\begin{eqnarray}
\nonumber&&\gamma_{12}^0=\frac{k_1+k_2-k_3-k_4}{2}\;,\quad \gamma_{13}^0=\frac{k_1+k_3-k_2-k_4}{2}\;, \\
&&\gamma_{34}^0=\gamma_{24}^0=0\;, \quad\gamma_{14}^0=k_4-L\;, \quad\gamma_{23}^0=\frac{k_4+k_2+k_3-k_1}{2}-L\;.
\end{eqnarray}
From the above definition, it follows that $\mathcal{G}^{I_1I_2I_3I_4}_{k_1k_2k_3k_4}(z,\bar{z};\alpha,\bar{\alpha})$ is a polynomial in $\alpha$ and $\bar{\alpha}$ with the same degree $L$. 

The cross ratios $z$, $\bar{z}$, $\alpha$, $\bar{\alpha}$ are related in a simple way to the cross ratios that appear in four-point correlators of SCFT$_{d\geq 3}$ with R-symmetry group $SO(d'\geq 5)$
\begin{equation}
U=\frac{x_{12}^2x_{34}^2}{x_{13}^2x_{24}^2}=z\bar{z}\;,\quad V=\frac{x_{14}^2x_{23}^2}{x_{13}^2x_{24}^2}=(1-z)(1-\bar{z})\;,
\end{equation}
\begin{equation}
\sigma=\frac{t_{13}t_{24}}{t_{12}t_{34}}=\alpha\bar{\alpha}\;,\quad \tau=\frac{t_{14}t_{23}}{t_{12}t_{34}}=(1-\alpha)(1-\bar{\alpha})\;.
\end{equation}
Here $t_\mu$ is a $d'$-dimensional null vector  satisfying $t^\mu t_\mu=0$ and $t_{ij}\equiv t_i^\mu t_{j \mu}$. When $d'=4$, we can construct a 4-dimensional null vector from the spinors,  $t^\mu\equiv\sigma^\mu_{\alpha\dot{\alpha}}v^\alpha\bar{v}^{\dot{\alpha}}$. Note that the two sets of cross ratios are inequivalent. Expressing $z$, $\bar{z}$ (or $\alpha$, $\bar{\alpha}$) in terms of $U$, $V$ (or $\sigma$, $\tau$) is generally ambiguous due to the appearance of square-roots when solving the quadratic equations\footnote{However if the function depends on $z$, $\bar{z}$ or $\alpha$, $\bar{\alpha}$ symmetrically, there is no such ambiguity.}.
In two dimensions, the four-point function $\mathcal{G}^{I_1I_2I_3I_4}_{k_1k_2k_3k_4}(z,\bar{z};\alpha,\bar{\alpha})$ is only invariant under simultaneously exchanging $(z,\alpha)\leftrightarrow(\bar{z},\bar{\alpha})$. This is in contrast to higher dimensions where the four-point function is always invariant under exchanging a single pair of cross ratios  $z\leftrightarrow\bar{z}$ and $\alpha\leftrightarrow\bar{\alpha}$. That is because in theories with $d\geq3$, $d'\geq5$ the four-point functions are functions of $x_{ij}^2$ and therefore can be written in terms of $U$ and $V$ (and similarly for the R-symmetry). On the other hand, in two dimensions there is one more structure, namely $\epsilon_{\mu\nu} x_{ij}^\mu x_{kl}^\nu$ and therefore the four-point function cannot be written as a function of $U$ and $V$ alone. This is the essential new feature of SCFT$_2$ compared to higher dimensions, which necessary requires to work with $(z,\bar{z};\alpha,\bar{\alpha})$.

So far we have only used the bosonic part of the global superconformal group. The fermionic generators impose extra constraints as the superconformal Ward identities\footnote{We have derived the superconformal Ward identities using two different methods. The first one uses the analytic superspace formalism, and is parallel to the analysis in \cite{Dolan:2004mu}. The second method uses a chiral algebra twist \cite{Beem:2013sza} on one of the $\mathfrak{psu}(1,1|2)$ subalgebra of the small $\mathcal{N}=4$ superconformal algebra. The second method is conceptually more interesting, and we will elaborate on it further in Appendix \ref{chiralalgebra}.}
\begin{eqnarray}
&&(\alpha \partial_\alpha-z\partial_z)\mathcal{G}^{I_1I_2I_3I_4}_{k_1k_2k_3k_4}\big|_{\alpha=1/z}=0\;,\\\label{scfwardid}
&&(\bar{\alpha} \partial_{\bar{\alpha}}-\bar{z}\partial_{\bar{z}})\mathcal{G}^{I_1I_2I_3I_4}_{k_1k_2k_3k_4}\big|_{\bar{\alpha}=1/\bar{z}}=0\;.
\end{eqnarray}
These identities can be solved as follows
\begin{equation}\label{scfwardidsol}
\mathcal{G}^{I_1I_2I_3I_4}_{k_1k_2k_3k_4}=\mathcal{G}^{I_1I_2I_3I_4}_{0,k_1k_2k_3k_4}+(1-z\alpha)(1-\bar{z}\bar{\alpha})\mathcal{H}^{I_1I_2I_3I_4}_{k_1k_2k_3k_4}
\end{equation}
where $\mathcal{G}^{I_1I_2I_3I_4}_{0,k_1k_2k_3k_4}(z,\bar{z};\alpha,\bar{\alpha})$ is a special solution which upon twisting becomes purely holomorphic (anti-holomorphic)
\begin{eqnarray}
\nonumber\mathcal{G}^{I_1I_2I_3I_4}_{0,k_1k_2k_3k_4}(z,\bar{z};\alpha,\bar{\alpha}=1/\bar{z})&=&f(z,\alpha)\;,\\  
\mathcal{G}^{I_1I_2I_3I_4}_{0,k_1k_2k_3k_4}(z,\bar{z};\alpha=1/z,\bar{\alpha})&=&f(\bar{z},\bar{\alpha})\;. 
\end{eqnarray}
The function $f(z,\alpha)$ can be further shown to be protected by non-renormalization theorems, by using the argument of \cite{Baggio:2012rr}. The function $\mathcal{H}^{I_1I_2I_3I_4}_{k_1k_2k_3k_4}(z,\bar{z};\alpha,\bar{\alpha})$ encodes unprotected dynamical information, and because of the prefactor $(1-z\alpha)(1-\bar{z}\bar{\alpha})$, $\mathcal{H}^{I_1I_2I_3I_4}_{k_1k_2k_3k_4}$ is a polynomial in both $\alpha$ and $\bar{\alpha}$ with the reduced degrees $L-1$.

\section{Position space}\label{Secpositionspace}
In this section, we develop a concrete position space method to compute holographic four-point functions. The method is very similar to the one used in \cite{Rastelli:2016nze,Rastelli:2017udc} for $AdS_5\times S^5$  and in \cite{Rastelli:2017ymc} for $AdS_7\times S^4$ . However, it has also new important ingredients due to the unique features of $AdS_3$ space. In Section \ref{reviewsugra} we review some elements of the 6d $(2,0)$ supergravity coupled to tensor multiplets, compactified on $AdS_3\times S^3$. Then, in Section \ref{posialgorithm} we outline the position space algorithm and in Section \ref{posikeq1} we compute the four-point function of the lowest-weight operator. Finally, in Section \ref{posihigherk} we apply the method to the four-point functions of higher weights.

\subsection{A brief review of $AdS_3\times S^3$ supergravity}\label{reviewsugra}
The near horizon limit of $Q_1$ D1-branes and $Q_5$ D5-branes wrapping a $K3$ surface is described by IIB supergravity in $AdS_3\times S^3\times K3$ when $Q_5 Q_1\gg1$. As in this limit the size of $K3$ is much smaller than the size of $S^3$, we can reduce IIB supergravity on K3 and get 6d $(2,0)$ supergravity coupled to $n=21$ anti-self-dual tensor multiplets. Then,  further compactification of the theory on $S^3$ gives the Kaluza-Klein spectrum \cite{Deger:1998nm,deBoer:1998kjm,deBoer:1998us,Arutyunov:2000by}  summarized in the tables \ref{tablegravtion}, \ref{tablescalarG},  \ref{tablescalarT} below.

\begin{table}[h]
\begin{center}\begin{tabular}{|c|c|c|c|c|c|c|}\hline KK mode & $h$ & $j$ & $\bar{h}$ & $\bar{j}$ &  spin $\ell$ & $R_{SO(21)}$ \\\hline $\varphi_{\mu\nu}^+$ & $\frac{l}{2}+2$ & $\frac{l}{2}$ & $\frac{l}{2}$ & $\frac{l}{2}$ & 2 & $\mathbf{1}$ \\\hline $V_\mu^+$ & $\frac{l}{2}+1$ &  $\frac{l}{2}+1$ & $\frac{l}{2}$ & $\frac{l}{2}$ & 1 & $\mathbf{1}$ \\\hline $W_\mu^+$ & $\frac{l}{2}+2$ & $\frac{l}{2}$ & $\frac{l}{2}+1$ & $\frac{l}{2}-1$ & 1 & $\mathbf{1}$ \\\hline  $\rho^+$ & $\frac{l}{2}+1$ & $\frac{l}{2}+1$ & $\frac{l}{2}+1$ & $\frac{l}{2}-1$ & 0 & $\mathbf{1}$ \\\hline \end{tabular} \caption{Kaluza Klein modes from the spin-2 multiplets $\Gamma_l$. We have kept only the relevant bosonic field modes which are singlets under the outer automorphism group $SO(4)_{out}$. Note that the fields $\varphi_{\mu\nu}^+$, $V_\mu^+$, $W_\mu^+$, $\rho^+$ are also accompanied by $\varphi_{\mu\nu}^-$, $V_\mu^-$, $W_\mu^-$, $\rho^-$ as required by parity. The quantum numbers $h$, $\bar{h}$, $j$, $\bar{j}$ of the $-$ fields are obtained from the quantum numbers of the $+$ fields by interchanging left and right. The number $l$ labels the Kaluza-Klein levels and $l=0,1,2,\ldots$. }\label{tablegravtion}
\end{center}
\end{table}

\begin{table}[h]
\begin{center}\begin{tabular}{|c|c|c|c|c|c|c|}\hline KK mode & $h$ & $j$ & $\bar{h}$ & $\bar{j}$ & spin $\ell$ & $R_{SO(21)}$ \\\hline $Y_\mu^+$ & $\frac{l}{2}+2$ & $\frac{l}{2}$ & $\frac{l}{2}+1$ & $\frac{l}{2}+1$ & 1 & $\mathbf{1}$ \\\hline $\sigma$ & $\frac{l}{2}+1$ &  $\frac{l}{2}+1$ & $\frac{l}{2}+1$ & $\frac{l}{2}+1$ & 0 & $\mathbf{1}$ \\\hline $\tau$ & $\frac{l}{2}+2$ & $\frac{l}{2}$ & $\frac{l}{2}+2$ & $\frac{l}{2}$ & 0 & $\mathbf{1}$ \\\hline $Y_\mu^-$ & $\frac{l}{2}+1$ & $\frac{l}{2}+1$ & $\frac{l}{2}+2$ & $\frac{l}{2}$ & 1 & $\mathbf{1}$ \\\hline \end{tabular} \caption{Kaluza-Klein modes from the spin-1 multiplet $\Sigma_l$, $l=0,1,\ldots$.}\label{tablescalarG}
\end{center}
\end{table}

\begin{table}[h]
\begin{center}\begin{tabular}{|c|c|c|c|c|c|c|}\hline KK mode & $h$ & $j$ & $\bar{h}$ & $\bar{j}$ & spin $\ell$ & $R_{SO(21)}$ \\\hline $Z_\mu^{+,I}$ & $\frac{l}{2}+2$ & $\frac{l}{2}$ & $\frac{l}{2}+1$ & $\frac{l}{2}+1$ & 1 & $\mathbf{21}$ \\\hline $s^I$ & $\frac{l}{2}+1$ &  $\frac{l}{2}+1$ & $\frac{l}{2}+1$ & $\frac{l}{2}+1$ & 0 & $\mathbf{21}$ \\\hline $t^I$ & $\frac{l}{2}+2$ & $\frac{l}{2}$ & $\frac{l}{2}+2$ & $\frac{l}{2}$ & 0 & $\mathbf{21}$ \\\hline $Z_\mu^{-,I}$ & $\frac{l}{2}+1$ & $\frac{l}{2}+1$ & $\frac{l}{2}+2$ & $\frac{l}{2}$ & 1 & $\mathbf{21}$ \\\hline \end{tabular} \caption{Kaluza-Klein modes from the spin-1 multiplet $\Theta_l^I$, $l=-1,0,1,\ldots$.}\label{tablescalarT}
\end{center}
\end{table}
The spectrum is organized into superconformal multiplets which come in three infinite Kaluza-Klein towers $\Gamma_l$, $\Sigma_l$ and $\Theta_l^I$. In the tables, $h$, $\bar{h}$ are the $SL(2)_L$, $SL(2)_R$ spins, and $j$, $\bar{j}$ are the $SU(2)_L$ and $SU(2)_R$ spins.  When the R-symmetry quantum numbers are negative, the corresponding field does not exist. We have only kept the fields that are singlets under the outer automorphism group $SO(4)_{out}$ because in this work we focus only on four-point functions of operators which are singlets. The superconformal primary of the multiplets $\Gamma_l$ which contains the spin-2 fields is the (massive) graviphoton field $V_\mu$. The lowest KK multiplet with $l=0$ is ultra-short: it contains only the non-dynamical massless graviton and graviphoton. The superconformal primaries of the spin-1 multiplets $\Sigma_l$ and $\Theta_l^I$ are the scalar fields $\sigma$ and $s^I$ respectively. These two multiplets have the same $SO(2,2)$ and $SO(4)_R$ quantum numbers, but $\Theta_l^I$ transform in the vector representation of $SO(21)$ while $\Sigma_l$ are singlets. In terms of 6d fields, $\Sigma_l$ is made of fields from 6d $(2,0)$ supergravity and $\Theta_l^I$ comes from the anti-self-dual tensors.
  Moreover, the minimal allowed value for $\Theta_l^I$ is $l=-1$, and the corresponding super primary has conformal dimension $\Delta=1$. The top component (not shown in the table) is an exactly marginal operator and transforms as a vector under $SO(4)_{out}$. By contrast, $\Sigma_l$ with $l=-1$ is pure-gauge and does not exist in the spectrum \cite{Deger:1998nm}. 

The cubic couplings of the Kaluza-Klein modes were obtained in \cite{Arutyunov:2000by}. The cubic couplings satisfy the R-symmetry selection rule, and vanish when they are extremal. The quartic and higher-oder vertices have not been worked out in the literature. Moreover, \cite{Deger:1998nm,Arutyunov:2000by} showed that the vector fields are described by two Proca-Chern-Simons vector fields supplemented by a first-order constraint. The vector fields couple to the currents made out of scalar fields both electrically and magnetically. We will show in Appendix \ref{firstordervectors} that the constraint can be solved in terms of three massive Chern-Simons fields which satisfy first-order equation of motions. After proper field redefinition, all the couplings of the vector fields with currents become electric.

\subsection{The position space algorithm}\label{posialgorithm}
We are now ready to formulate the position space method. We start with an ansatz for the four-point function which includes all the possible exchange and contact Witten diagrams
\begin{equation}
\begin{split}
\mathcal{A}^{I_1I_2I_3I_4}={}&\delta^{I_1I_2}\delta^{I_3I_4}\mathcal{A}^{\text{s-exch}}+\delta^{I_1I_4}\delta^{I_2I_3}\mathcal{A}^{\text{t-exch}}+\delta^{I_1I_3}\delta^{I_2I_4}\mathcal{A}^{\text{u-exch}}\\
{}&+\delta^{I_1I_2}\delta^{I_3I_4}\mathcal{A}^{\text{s-con}}+\delta^{I_1I_4}\delta^{I_2I_3}\mathcal{A}^{\text{t-con}}+\delta^{I_1I_3}\delta^{I_2I_4}\mathcal{A}^{\text{u-con}}\;.
\end{split}
\end{equation}
The exchange Witten diagrams are subject to the R-symmetry selection rule and the requirement that the cubic coupling is non-extremal. In addition, the contact Witten diagrams should contain no more than two derivatives. This condition comes from the consistency with the flat space limit in which the theory contains only two derivatives. The next step is to evaluate all the diagrams in the ansatz. Compared to the $AdS_5\times S^5$ case, there are two new kinds of Witten diagrams. The first is the exchange diagram of twist-zero fields which are the massless Chern-Simons and the graviton field. The standard method of \cite{DHoker:1999mqo} is not applicable for these diagrams.\footnote{If one naively applies the method of \cite{DHoker:1999mqo}, one finds the answer is divergent. The unphysical divergence is associated with dropping certain boundary terms in the analysis which is not allowed for $d=2$.} We instead evaluate them by solving second order differential equations with appropriate boundary conditions. These differential equations follow from the simple fact that the two-particle quadratic conformal Casimir is the same as the wave equation in the bulk, which collapses the exchange diagram into a contact diagram  when acting on the bulk-to-bulk propagator.\footnote{This fact was also recently exploited in \cite{Zhou:2018sfz} to obtain the conformal block decomposition coefficients of exchange Witten diagrams and conformal partial waves in the crossed channel.} We leave the details of this method to Appendix \ref{AdS3Wdiagrams}. The second type of new diagram is the exchange diagrams which involve massive Chern-Simons vector fields. This type of diagrams can be evaluated using the method of \cite{DHoker:1999mqo} with slight modifications. All in all, all the Witten diagrams can be evaluated in terms of a finite sum of contact diagrams ($\bar{D}$-functions).
To impose the superconformal Ward identities (\ref{scfwardid}), we exploit the fact that $\bar{D}$-functions can be uniquely decomposed as
\begin{equation}
\bar{D}_{\Delta_1\Delta_2\Delta_3\Delta_4}=R_\Phi(z,\bar{z}) \Phi(z,\bar{z})+R_V(z,\bar{z})\log(1-z)(1-\bar{z})+R_U(z,\bar{z})\log(z\bar{z})+R_1(z,\bar{z})
\end{equation}
where $\Phi(z,\bar{z})=\bar{D}_{1111}$ is the scalar box diagram, and $R_{\Phi,U,V,1}$ are rational functions of $z$ and $\bar{z}$. It is clear that the supergravity ansatz admits a similar decomposition. By further using the recursion relation \cite{Eden:2000bk}
\begin{eqnarray}
\nonumber&&\partial_z\Phi=\frac{\Phi}{\bar{z}-z}+\frac{\log (1-z)(1-\bar{z})}{z(\bar{z}-z)}+\frac{\log(z\bar{z})}{(z-1)(z-\bar{z})}\;,\\
&&\partial_{\bar{z}}\Phi=\frac{\Phi}{z-\bar{z}}+\frac{\log (1-z)(1-\bar{z})}{\bar{z}(z-\bar{z})}+\frac{\log(z\bar{z})}{(\bar{z}-1)(\bar{z}-z)}\;,
\end{eqnarray}
we can similarly decompose the left side of the superconformal Ward identity (\ref{scfwardid}) into this basis. The rational coefficient functions $R_{{\rm Ward},X}^{I_1I_2I_3I_3}(z,\bar{z},\alpha,\bar{\alpha})$ with $X=\Phi,U,V,1$ are required to vanish by (\ref{scfwardid}), giving rise to a set of linear equations for the unknown coefficients in the ansatz. In contrast to the $AdS_5\times S^5$ and $AdS_7\times S^4$ cases, solving superconformal Ward identities in general does not uniquely fix all the relative coefficients. We will see that all the coefficients in $\mathcal{A}_{\rm con}^{I_1I_2I_3I_4}$ parameterizing the quartic vertices are fixed in terms of the coefficients in the exchange part of the ansatz. The remaining unsolved coefficients can be fixed by comparing with the known supergravity cubic couplings.

\subsection{The lowest-weight four-point function}\label{posikeq1}

We now apply the above method to the simplest four-point correlator with $k_i=k=1$. The cubic coupling selection rules dictate that only the non-dynamical graviton and Chern-Simons gauge field can be exchanged. Therefore we have the following ansatz for the exchange part of the four-point function
\begin{equation}
\mathcal{A}^{\rm s-exch}=\lambda_{gr} \underbrace{Y_0\bar{Y}_0\mathcal{W}_{gr}}_{\varphi_{l=0,\mu\nu}}+\lambda_{CS}(\underbrace{Y_1\bar{Y}_0\mathcal{W}^{CS,1,+}}_{V_{l=0,\mu}^+}+\underbrace{\bar{Y}_1Y_0\mathcal{W}^{CS,1,-}}_{V_{l=0,\mu}^-}),
\end{equation}
where $Y_m$ and $\bar{Y}_{\bar{m}}$ are the $SU(2)_L$ and $SU(2)_R$ R-symmetry polynomials
\begin{equation}
Y_m=P_m(1-2\alpha)\;,\quad \bar{Y}_{\bar{m}}=P_{\bar{m}}(1-2\bar{\alpha})\;,
\end{equation}
associated with exchanging the representation $(j,\bar{j})=(m,\bar{m})$. The function $\mathcal{W}_{gr}$ is the exchange Witten diagram of the non-dynamical graviton, and is worked out in Appendix \ref{AdS3Wdiagrams} to be
\begin{equation}
\mathcal{W}_{gr}=\frac{\pi}{2}(2+U(U-V-1)\bar{D}_{2211})\;.
\end{equation} 
Similarly, $\mathcal{W}^{CS,1,+}$, $\mathcal{W}^{CS,1,-}$ are contributions of the Witten diagrams with the massless Chern-Simons gauge field  $V_{l=0,\mu}^+$, $V_{l=0,\mu}^-$ being exchanged. They are given by
\begin{equation}
\mathcal{W}^{CS,1,\pm}=\frac{\pi}{2}\left(\mp (z-\bar{z})U\bar{D}_{2211}+\log V\right)\;.
\end{equation} 
The ansatz for $\mathcal{A}^{\rm t-exch}$, $\mathcal{A}^{\rm u-exch}$ are obtained from $\mathcal{A}^{\rm s-exch}$ using crossing symmetry. The contact part of the ansatz takes the from 
\begin{equation}
\begin{split}
\mathcal{A}^{\text{s-con}}={}&\sum c^{(0)}_{ab}\sigma^a\tau^b U \bar{D}_{1111}+\sum c^{(1,s)}_{ab}\sigma^a\tau^b U^2 \bar{D}_{2211}+\sum c^{(1,t)}_{ab}\sigma^a\tau^b U \bar{D}_{2112}\\
{}&+\sum c^{(1,u)}_{ab}\sigma^a\tau^b U \bar{D}_{2121}\; ,
\end{split}
\end{equation}
where the sum is restricted by $0\leq a,b,a+b\leq 1$. Note that no individual $\alpha$, $\bar{\alpha}$ appears in the ansatz because the four-point function is parity even under separate exchange of $z\leftrightarrow\bar{z}$ and $\alpha\leftrightarrow\bar{\alpha}$. The contribution of the contact diagrams in the other two channels $\mathcal{A}^{\text{t-con}}$, $\mathcal{A}^{\text{u-con}}$ are obtained from $\mathcal{A}^{\text{s-con}}$ using crossing symmetry. 

Plugging this ansatz into the superconformal Ward identities (\ref{scfwardid}), we find that 
\begin{equation}
\lambda_{CS}=\frac{1}{2}\lambda_{gr}\;,
\end{equation}
and all the contact term coefficients are solved in terms of $\lambda_{gr}$. Therefore the four-point function is fixed up to an overall coefficient. Rewriting the solution in the form of (\ref{scfwardidsol}) we find that 
\begin{equation}\label{G0keq1posi}
\mathcal{G}^{I_1I_2I_3I_4}_{0,1111}=\frac{\pi \lambda_g}{V}(V\delta^{I_1I_2}\delta^{I_3I_4}+U\tau \delta^{I_1I_4}\delta^{I_2I_3}+UV\sigma \delta^{I_1I_3}\delta^{I_2I_4})\;,
\end{equation}
\begin{equation}\label{Hkeq1posi}
\mathcal{H}^{I_1I_2I_3I_4}_{1111}=-\frac{\pi \lambda_g}{V}(\delta^{I_1I_2}\delta^{I_3I_4}V\bar{D}_{1122}+ \delta^{I_1I_4}\delta^{I_2I_3}U\bar{D}_{2112}+ \delta^{I_1I_3}\delta^{I_2I_4}UV\bar{D}_{1212})\;,
\end{equation}
reproducing the result of \cite{Giusto:2018ovt}.

\subsection{Higher-weight four-point functions}\label{posihigherk}
Let us move on to the next simplest correlator with $k=2$. Our ansatz for the singular part of the four-point function is 
\begin{equation}
\begin{split}
\mathcal{A}^{\rm s-exch}={}&\lambda_{gr}^{(0)} \underbrace{Y_0\bar{Y}_0\mathcal{W}_{gr}}_{\varphi_{l=0,\mu\nu}}+\lambda_{CS}^{(0)}(\underbrace{Y_1\bar{Y}_0\mathcal{W}^{CS,1,+}}_{V_{l=0,\mu}^+}+\underbrace{\bar{Y}_1Y_0\mathcal{W}^{CS,1,-}}_{V_{l=0,\mu}^-})\\
+{}&\lambda_{\varphi}^{(2)}\underbrace{Y_1\bar{Y}_1\mathcal{W}_{mgr,4}}_{\varphi_{l=2,\mu\nu}}+\lambda_{V}^{(2)}(\underbrace{Y_2\bar{Y}_1\mathcal{W}^{CS,3,+}}_{V_{l=2,\mu}^+}+\underbrace{\bar{Y}_2Y_1\mathcal{W}^{CS,3,-}}_{V_{l=2,\mu}^-})\\
+{}&\lambda_\sigma^{(0)} \underbrace{Y_1\bar{Y}_1\mathcal{W}_{sc,2}}_{\sigma_{l=0}}+\lambda_{Y}^{(0)}(\underbrace{Y_0\bar{Y}_1\mathcal{W}^{CS,3,+}}_{Y_{l=0,\mu}^-}+\underbrace{\bar{Y}_0Y_1\mathcal{W}^{CS,3,-}}_{Y_{l=0,\mu}^+})\;.
\end{split}
\end{equation}
Here $\mathcal{W}_{mgr,4}$ is the exchange diagram of a massive graviton of dimension 4 and $\mathcal{W}_{sc,2}$ is a scalar exchange diagram of dimension 2. Both diagrams can be computed using the method of \cite{DHoker:1999mqo}. The contact part ansatz $\mathcal{A}^{\rm s-con}$ contains zero and two-derivative contact Witten diagrams, and is a polynomial in $\sigma$ and $\tau$ of degree 2. Solving the superconformal Ward identities uniquely fixes the coefficients in  $\mathcal{A}^{\rm s-con}$ in terms of the coefficients appearing in $\mathcal{A}^{\rm s-exch}$. Moreover, the coefficients of the exchange contributions of fields belonging to the same multiplet are fixed up to an overall normalization. The remaining unfixed relative coefficients corresponding to different multiplets can be fixed using the cubic vertices worked out in \cite{Arutyunov:2000by}. The final solution can be rewritten in the form of (\ref{scfwardidsol}) as
\begin{equation}\label{G0keq2posi}
\mathcal{G}^{I_1I_2I_3I_4}_{0,2222}=\frac{\pi \lambda_g}{4V^2}(V^2\delta^{I_1I_2}\delta^{I_3I_4}+U^2\tau^2 \delta^{I_1I_4}\delta^{I_2I_3}+U^2V^2\sigma^2 \delta^{I_1I_3}\delta^{I_2I_4})\;,
\end{equation}
\begin{equation}\label{H0keq2posi}
\begin{split}
\mathcal{H}^{I_1I_2I_3I_4}_{2222}={}&-\frac{\pi \lambda_g}{4V}\left(\delta^{I_1I_2}\delta^{I_3I_4}V(U\bar{D}_{2233}+\bar{D}_{1133})+\delta^{I_1I_4}\delta^{I_2I_3}U^2\bar{D}_{3223}+\delta^{I_1I_3}\delta^{I_2I_4}U^2V\bar{D}_{2323}\right)\\
{}&+crossing\;.
\end{split}
\end{equation}
The case of higher-weight correlators with $k>2$ is completely analogous to the above example with $k=2$. We have also applied this method to obtain four-point correlators for two more examples with $k=3$ and $k=4$. We will refrain from writing down the explicit results, since in the Section \ref{generatingfunction} we will present a much more compact way of writing these correlators. 

\section{Mellin space}\label{SecMellinspace}
The position space method described in Section \ref{posialgorithm} offers a concrete way to compute four-point functions with as little input from supergravity. However, the results in position space do not look particularly illuminating. In this section, we look at the problem from a different perspective using the Mellin representation formalism \cite{Mack:2009mi,Penedones:2010ue,Paulos:2011ie,Fitzpatrick:2011ia}, which offers new intuition to the problem.  The Mellin representation formalism was demonstrated to be the most natural language for describing holographic correlators, making manifest their scattering amplitude nature. This formalism unfortunately becomes ill-defined in one dimension due to the nonlinear dependence of the cross ratios\footnote{For $d\geq2$ there are two independent conformal cross ratios $U$, $V$ in four-point functions, while for $d=1$ there is only one independent cross ratio $z$. More generally, there are $n(n-3)/2$ independent cross ratios for a scalar $n$-point function if $n\leq d+2$. When $n>d+2$ the cross ratios have relations.}. Because the superconformal symmetry forces the chiral cross ratios $z$, $\bar{z}$ to appear in the 2d one-half BPS correlator $\mathcal{G}^{I_1I_2I_3I_4}_{k_1k_2k_3k_4}$, one may wonder if the Mellin representation will be particularly useful. Nevertheless, by restricting our attention to certain components of the four-point function we can  argue that the Mellin representation is still a good language. In particular, the Mellin representation formalism allows us to easily bootstrap the $k_i=k=1$ correlator as we demonstrate below.

For $k=1$, $\mathcal{H}^{I_1I_2I_3I_4}_{k=1}$ has no dependence on $\alpha$ and $\bar{\alpha}$. The symmetry under $z\leftrightarrow\bar{z}$, $\alpha\leftrightarrow\bar{\alpha}$ implies that $\mathcal{H}^{I_1I_2I_3I_4}_{k=1}$ is a symmetric function of  $z$, $\bar{z}$ and can be unambiguously expressed in terms of $U$, $V$. We therefore have the following inverse Mellin representation
\begin{equation}\label{MellinHkeq1}
\mathcal{H}^{I_1I_2I_3I_4}_{k=1}=\int_{-i\infty}^{i\infty}\frac{ds}{2}\frac{dt}{2}U^{\frac{s}{2}}V^{\frac{t}{2}-1}\widetilde{\mathcal{M}}^{I_1I_2I_3I_4}_{k=1}(s,t)\Gamma^2(\frac{2-s}{2})\Gamma^2(\frac{2-t}{2})\Gamma^2(\frac{2-\tilde{u}}{2})\; ,
\end{equation}
where $\tilde{u}=2-s-t$. We assume that $\mathcal{G}^{I_1I_2I_3I_4}_{0,k=1}$ is a rational function (this was justified by the previous position space calculation) and therefore does not contribute to the Mellin amplitude\footnote{The rational terms are generated from regularization effects when the integration contours are pinched. See \cite{Rastelli:2017udc} for details.}. It then follows that the following components of $\mathcal{G}^{I_1I_2I_3I_4}_{k=1}$ with R-symmetry factors 
\begin{equation}
\mathcal{P}_1\equiv1\;,\;\; \mathcal{P}_2\equiv\frac{\alpha+\bar{\alpha}}{2}\;,\;\; \mathcal{P}_3\equiv\alpha\bar{\alpha}=\tau\;,
\end{equation}
also have a well-defined Mellin representation. By rewriting the superconformal factor $(1-z\alpha)(1-\bar{z}\bar{\alpha})$ as
\begin{equation}
1+\frac{\alpha+\bar{\alpha}}{2}(V-U-1)+\alpha\bar{\alpha} U+\frac{\alpha-\bar{\alpha}}{2}(z-\bar{z})\;,
\end{equation}
we read off the three R-symmetry components of $\mathcal{G}^{I_1I_2I_3I_4}_{k=1}$ 
\begin{eqnarray}
&&\mathcal{P}_1\;:\quad\quad\mathcal{G}^{I_1I_2I_3I_4}_{k=1,\mathcal{I}=1}\equiv\mathcal{H}^{I_1I_2I_3I_4}_{k=1}\;,\\
&&\mathcal{P}_2\;:\quad\quad\mathcal{G}^{I_1I_2I_3I_4}_{k=1,\mathcal{I}=2}\equiv(V-U-1)\mathcal{H}^{I_1I_2I_3I_4}_{k=1}\;,\\
&&\mathcal{P}_3\;:\quad\quad\mathcal{G}^{I_1I_2I_3I_4}_{k=1,\mathcal{I}=3}=U\mathcal{H}^{I_1I_2I_3I_4}_{k=1}\;.
\end{eqnarray}
They can be expressed in the same form as (\ref{MellinHkeq1})
\begin{equation}
\mathcal{G}^{I_1I_2I_3I_4}_{k=1, \mathcal{I}}=\int_{-i\infty}^{i\infty}\frac{ds}{2}\frac{dt}{2}U^{\frac{s}{2}}V^{\frac{t}{2}-1}\mathcal{M}^{I_1I_2I_3I_4}_{k=1, \mathcal{I}}(s,t)\Gamma^2(\frac{2-s}{2})\Gamma^2(\frac{2-t}{2})\Gamma^2(\frac{2-u}{2})\;,
\end{equation}
with $ u=4-s-t$, by absorbing the multiplicative monomials $U^mV^n$ via shifting $s$ and $t$. The monomials then become difference operators which act as 
\begin{equation}
\widehat{U^mV^n}\circ \widetilde{\mathcal{M}}^{I_1I_2I_3I_4}_{k=1}(s,t)=\widetilde{\mathcal{M}}^{I_1I_2I_3I_4}_{k=1}(s-2m,t-2n)\left(\frac{2-s}{2}\right)^2_m\left(\frac{2-t}{2}\right)^2_n\left(\frac{s+t-2}{2}\right)^2_{1-m-n}\;.
\end{equation}
We are now ready to formulate a bootstrap problem for $\widetilde{\mathcal{M}}^{I_1I_2I_3I_4}_{k=1}$ by enumerating the extra constraints that should be satisfied by $\mathcal{M}^{I_1I_2I_3I_4}_{k=1, \mathcal{I}}$. 
\bigskip

1. {\bf Bose symmetry.} The Mellin amplitudes $\mathcal{M}^{I_1I_2I_3I_4}_{k=1, \mathcal{I}}$ are crossing symmetric. It is convenient to first make the flavor dependence more explicit
\begin{equation}
\mathcal{M}^{I_1I_2I_3I_4}_{k=1, \mathcal{I}}(s,t)=\delta^{I_1I_2}\delta^{I_3I_4}\mathcal{M}^{(s)}_{k=1, \mathcal{I}}(s,t)+\delta^{I_1I_4}\delta^{I_2I_3}\mathcal{M}^{(t)}_{k=1, \mathcal{I}}(s,t)+\delta^{I_1I_3}\delta^{I_2I_4}\mathcal{M}^{(u)}_{k=1, \mathcal{I}}(s,t)\;.
\end{equation}
Crossing symmetry then implies that
\begin{eqnarray}
\nonumber\left[\begin{array}{c}\mathcal{M}^{(t)}_{k=1, 1}(s,t) \\\mathcal{M}^{(t)}_{k=1, 2}(s,t) \\\mathcal{M}^{(t)}_{k=1, 3}(s,t)\end{array}\right]&=&\left[\begin{array}{ccc}1 & 0 & 0 \\-2 & 1 & 0 \\1 & -1 & 1\end{array}\right]\left[\begin{array}{c}\mathcal{M}^{(s)}_{k=1, 1}(t,s) \\\mathcal{M}^{(s)}_{k=1, 2}(t,s) \\\mathcal{M}^{(s)}_{k=1, 3}(t,s)\end{array}\right]\;,\\
\mathcal{M}^{(u)}_{k=1, \mathcal{I}}(s,t)&=&\mathcal{M}^{(s)}_{k=1, \mathcal{I}}(u,t)\;.
\end{eqnarray}

 2. {\bf Analytic structure.} The four-point function can be computed as a sum of Witten diagrams. The conformal block decomposition consists of only single-trace operators and double-trace operators. The twists of exchanged single-trace operators translate into the position of simple poles in the Mellin amplitudes, while the double-trace operators are already accounted by the Gamma function factors. From the supergravity spectrum we expect that $\mathcal{M}^{(s)}_{k=1, \mathcal{I}}$ contain only a simple pole at $s=0$ due to the s-channel exchange of the non-dynamical graviton and gauge field. Similarly, $\mathcal{M}^{(t)}_{k=1, \mathcal{I}}$ and $\mathcal{M}^{(u)}_{k=1, \mathcal{I}}$ have only simple poles at $t=0$ and $u=0$ respectively. Furthermore, the residue at each pole is a polynomial in the other Mandelstam variable.

3. {\bf Asymptotics.} $\mathcal{M}^{(a)}_{k=1, \mathcal{I}}$ should grow linearly at large values of the Mandelstam variables,
\begin{equation}
\mathcal{M}^{(a)}_{k=1, \mathcal{I}}(\beta s,\beta t)\sim O(\beta)\;,\quad \text{for}\; \beta\to\infty\;.
\end{equation}
This comes from the expectation that the Mellin amplitudes $\mathcal{M}^{(a)}_{k=1, \mathcal{I}}$ in the asymptotic regime should reproduce the flat space scattering amplitudes of tensors in 6d $(2,0)$ supergravity \cite{Penedones:2010ue}.
\bigskip

These conditions turn out to be constraining enough, and uniquely fix the Mellin amplitude up to an overall factor 
\begin{equation}\label{Mtildekeq1}
\widetilde{\mathcal{M}}^{I_1I_2I_3I_4}_{k=1}(s,t)\propto \frac{\delta^{I_1I_2}\delta^{I_3I_4}}{s}+\frac{\delta^{I_1I_4}\delta^{I_2I_3}}{t}+\frac{\delta^{I_1I_3}\delta^{I_2I_4}}{\tilde{u}}\;.
\end{equation}
Translating the result back into $\mathcal{H}^{I_1I_2I_3I_4}$, we find 
\begin{equation}\label{calH1111}
\mathcal{H}^{I_1I_2I_3I_4}\propto\delta^{I_1I_2}\delta^{I_3I_4}V^{-1}\bar{D}_{1122}+{\delta^{I_1I_4}\delta^{I_2I_3}}UV^{-1}\bar{D}_{2112}+\delta^{I_1I_3}\delta^{I_2I_4}UV^{-1}\bar{D}_{1212}\;,
\end{equation}
which reproduces our previous result (\ref{Hkeq1posi}) in position space. 

In fact the above arguments apply more generally to a class of four-point correlators with have the same extremality\footnote{Extremality $E$ is defined as $E=k_2+k_3+k_4-k_1$ where $k_1$ is the largest of all $k_i$.}, {\it e.g.}, $k_1=k_2=n$, $k_3=k_4=1$. For these near-extremal correlators, the auxiliary Mellin amplitudes $\widetilde{\mathcal{M}}^{I_1I_2I_3I_4}$ are uniquely determined by the bootstrap conditions up to an overall coefficient, and take the same form as (\ref{Mtildekeq1}) with shifted simple poles. 

Let us also make two comments about applying the Mellin space method to correlators with higher extremities. First of all, it is necessary to make the assumption that $\mathcal{H}^{I_1I_2I_3I_4}$ is even under $\alpha\leftrightarrow\bar{\alpha}$, or in other words, can be uniquely expressed in terms of $\sigma$ and $\tau$. This is needed such that $\mathcal{H}^{I_1I_2I_3I_4}$ can be unambiguously written in terms of $U$ and $V$, and therefore admits a well-defined Mellin representation. The R-symmetry structure of $\mathcal{H}^{I_1I_2I_3I_4}$ {\it a priori} can be more general. However the even parity of $\mathcal{H}^{I_1I_2I_3I_4}$ is observed in all examples computed using the position space method, and we believe is true in general. Second, the bootstrap conditions are not as constraining as the $AdS_5\times S_5$ case. We expect that $\widetilde{\mathcal{M}}^{I_1I_2I_3I_4}$ also takes the form of a sum of simple poles. However, the pole structure of the ansatz makes the condition on analytic structures weaker. In particular, the requirement that $\mathcal{M}^{I_1I_2I_3I_4}$ should have polynomial residues is now trivially satisfied due to the lack of simultaneous poles $\widetilde{\mathcal{M}}^{I_1I_2I_3I_4}$. Therefore, unlike the $AdS_5\times S_5$ case where the Mellin amplitude is fixed up to an overall factor, we cannot solve all the parameters in the ansatz for the Mellin amplitude. This parallels what we have observed in the position space method, and is an inevitable consequence of the fact that we have fewer supersymmetry for $AdS_3\times S^3\times K3$.

\section{Hidden symmetry and conjecture of all Kaluza-Klein four-point functions}\label{Sechiddensymm}
\subsection{Hidden conformal symmetry}
The general one-half BPS four-point functions of 4d $\mathcal{N}=4$ super Yang-Mills theory in the supergravity limit were obtained in \cite{Rastelli:2016nze,Rastelli:2017udc} by solving an algebraic bootstrap problem. The formula took a surprisingly simple form and therefore strongly suggested the existence of some elegant underlying structure. Recently, this was made precise by  \cite{Caron-Huot:2018kta} in terms of a conjectured 10d conformal symmetry. In terms of this symmetry, all one-half BPS four-point functions can be organized into one generating function, which is obtained by promoting the 4d distances in the lowest-weight correlator into 10d distances. Though a rigorous understanding is still lacking  regarding its origin, some intuitive arguments were given in \cite{Caron-Huot:2018kta} to motivate the existence of such a symmetry. We will enumerate below some of these arguments and we will see that many features are also shared by $AdS_3\times S^3$. 

First of all, the $AdS_5\times S^5$ background is conformally equivalent to the flat space $\mathbb{R}^{9,1}$. The $SO(10,2)$ symmetry can be interpreted as the conformal group in $\mathbb{R}^{9,1}$. The same statement can be made for the $AdS_3\times S^3$ background and the conformal group $SO(6,2)$.\footnote{More precisely, it requires that the AdS space should have the same radius as the sphere. This is true for $AdS_5\times S^5$ and $AdS_3\times S^3$, but is not true for, {\it e.g.}, $AdS_7\times S^4$ which is dual to 6d $(2,0)$ SCFTs.} Secondly, it was argued that the $AdS_5\times S^5$ auxiliary Mellin amplitude $\widetilde{\mathcal{M}}$ of \cite{Rastelli:2016nze,Rastelli:2017udc} 
\begin{equation}
\widetilde{\mathcal{M}}\sim \frac{1}{(s-2)(t-2)(\tilde{u}-2)}\; ,
\end{equation}
should be identified, in the large Mellin variable limit, with the $\frac{1}{stu}$ factor in the superamplitude of IIB supergravity in 10d flat space
\begin{equation}
\mathcal{A}^{\rm IIB}\sim G^{(10)}_N\delta^{16}(Q)\times \frac{1}{stu}\;.
\end{equation}
When divided by the dimensionless ``coupling'' $G_{N}^{(10)}\delta^{16}(Q)$, the amplitude $\frac{1}{stu}$ is conformal invariant in ten dimensions, {\it i.e.}, annihilated by the special conformal transformation generator\footnote{We use momentum conservation to solve $p_4$ in terms of $p_1$, $p_2$, $p_3$, and write
$\frac{1}{stu}=\frac{1}{(p_1\cdot p_2)(p_1\cdot p_3)(p_2\cdot p_3)}$.}
\begin{equation}
K_\mu=\sum_{i=1}^3\left(\frac{p_{i\mu}}{2}\frac{\partial}{\partial p_i^\nu}\frac{\partial}{\partial p_{i,\nu}}-p_i^\nu\frac{\partial}{\partial p_i^\nu}\frac{\partial}{\partial p_i^\mu}-\frac{d-2}{2}\frac{\partial}{\partial p_i^\mu}\right)\;.
\end{equation} 
For $AdS_3\times S^3$ we found a highly nontrivial analogy. By taking the asymptotic limit of the auxiliary Mellin amplitude $\widetilde{\mathcal{M}}^{I_1I_2I_3I_4}_{k=1}$, we find that we precisely reproduce the four tensor scattering amplitude in the theory of 6d (2,0) supergravity coupled to 21 Abelian tensor multiplets \cite{Lin:2015dsa}
\begin{equation}
\mathcal{A}^{\text{(2,0)}}\sim G_{N}^{(6)}\delta^{8}(Q)\left(\frac{\delta^{I_1I_2}\delta^{I_3I_4}}{s}+\frac{\delta^{I_1I_4}\delta^{I_2I_3}}{t}+\frac{\delta^{I_1I_3}\delta^{I_2I_4}}{u}\right)\;.
\end{equation}
After dividing by the dimensionless quantity $G_{N}^{(6)}\delta^{8}(Q)$, the amplitude is also conformally invariant in 6d. Finally, \cite{Caron-Huot:2018kta} also observed that the form of double-trace anomalous dimensions coincides with the partial wave decomposition coefficients of the 10d scattering amplitudes. Moreover the use of the 10d conformal block diagonalizes the mixing problem of double-trace operators. We have not investigated the counterparts of these problems in six dimensions, but it is likely that the questions will have similar answers. We hope to return to these questions in the future. 

\subsection{Conjecture of four-point functions with general weights}\label{generatingfunction}
Motivated by the above similarities, we propose that a hidden $SO(6,2)$ symmetry exists for $AdS_3\times S^3$, in the same sense of the $AdS_5\times S^5$ case. This symmetry will translate into a prescription for writing down a generating function for all one-half BPS four-point functions. 

Let us define from $\mathcal{H}^{I_1I_2I_3I_4}$ a crossing-symmetric function 
\begin{equation}
H_{k_1k_2k_3k_4}^{I_1I_2I_3I_4}=\mathbf{K} \left(\frac{t_{12}t_{34}}{x_{12}^2x_{34}^2}\right)^{-1}\frac{\mathcal{H}^{I_1I_2I_3I_4}_{k_1k_2k_3k_4}}{x_{12}^2x_{34}^2x_{13}^2x_{24}^2}\;.
\end{equation}
In particular, for $k_i=1$
\begin{equation}
H_{1111}^{I_1I_2I_3I_4}=\frac{\mathcal{H}^{I_1I_2I_3I_4}_{1111}}{x_{12}^2x_{34}^2x_{13}^2x_{24}^2}\equiv\frac{\mathcal{H}^{I_1I_2I_3I_4}_{k=1}}{x_{12}^2x_{34}^2x_{13}^2x_{24}^2} \; ,
\end{equation}
is a function of $x_{ij}^2$ only. Our main contention is that $H^{I_1I_2I_3I_4}_{1111}$ can be promoted into a {\it generating function} by doing a simple replacement in the arguments
\begin{equation}\label{genfunction}
\mathbf{H}(x_i,t_i)^{I_1I_2I_3I_4}\equiv H_{1111}^{I_1I_2I_3I_4}(x_{ij}^2+t_{ij})\;.
\end{equation}
All the functions $H_{k_1k_2k_3k_4}^{I_1I_2I_3I_4}$ with higher values of $k_i$ are obtained by first expanding $\mathbf{H}(x_i,t_i)$ in powers of $t_{ij}$ and then collecting all the possible monomials $\prod_{i<j}(t_{ij})^{\gamma_{ij}}$ that appear in $H_{k_1k_2k_3k_4}^{I_1I_2I_3I_4}$. For example, 
\begin{eqnarray}
&&\nonumber H_{11nn}^{I_1I_2I_3I_4}\propto t_{34}^{n-1}\bigg(\delta^{I_1I_2}\delta^{I_3I_4}\frac{D_{1,1,n+1,n+1}}{x_{12}^2}+{\delta^{I_1I_4}\delta^{I_2I_3}}\frac{D_{2,1,n,n+1}}{x_{23}^2}\\
&&\quad\quad\quad\quad+\delta^{I_1I_3}\delta^{I_2I_4}\frac{D_{1,2,n,n+1}}{x_{13}^2}\bigg)\;,\\
\nonumber &&H_{2222}^{I_1I_2I_3I_4}\propto t_{12}t_{34}\bigg(\delta^{I_1I_2}\delta^{I_3I_4}\left(\frac{D_{2233}}{x_{12}^2}+3\frac{D_{1133}}{x_{12}^4}\right)+\delta^{I_1I_3}\delta^{I_2I_4}\frac{D_{2323}}{x_{13}^2}+\delta^{I_1I_4}\delta^{I_2I_3}\frac{D_{3223}}{x_{23}^2}\bigg)\\
\nonumber&&\quad\quad+t_{14}t_{23}\bigg(\delta^{I_1I_4}\delta^{I_2I_3}\left(\frac{D_{3223}}{x_{23}^2}+3\frac{D_{3113}}{x_{23}^4}\right)+\delta^{I_1I_3}\delta^{I_2I_4}\frac{D_{2323}}{x_{13}^2}+\delta^{I_1I_2}\delta^{I_3I_4}\frac{D_{2233}}{x_{12}^2}\bigg)\\
&&\quad\quad+t_{13}t_{24}\bigg(\delta^{I_1I_3}\delta^{I_2I_4}\left(\frac{D_{2323}}{x_{13}^2}+3\frac{D_{1313}}{x_{13}^4}\right)+\delta^{I_1I_2}\delta^{I_3I_4}\frac{D_{2233}}{x_{12}^2}+\delta^{I_1I_4}\delta^{I_2I_3}\frac{D_{3223}}{x_{23}^2}\bigg)\;.
\end{eqnarray}
The last expression is nothing but (\ref{H0keq2posi}). We have also checked that expanding the generating function gives $H_{kkkk}^{I_1I_2I_3I_4}$ which agree completely with our position space calculations for $k=3,4$.

We also conjecture that $\mathcal{G}_{0,k_1k_2k_3k_4}^{I_1I_2I_3I_4}$ takes the form of generalized free fields correlator,  plus the contributions due to the mixing of the single-trace operator with double-trace operators in the external operators \cite{Arutyunov:2000ima,Taylor:2007hs}. This is based on the explicit examples with $k=1,2,3,4$ which we have computed using the position space method. In principle, given $H_{k_1k_2k_3k_4}^{I_1I_2I_3I_4}$ the function $\mathcal{G}_{0,k_1k_2k_3k_4}^{I_1I_2I_3I_4}$ can be fixed by requiring consistency with the conformal block decomposition. For example, if we consider four-point functions with $\{k_1,k_2,k_3,k_4\}=\{n,n,1,1\}$, consistency with the single-trace conformal blocks in the decomposition fixes $\mathcal{G}_{0,nn11}^{I_1I_2I_3I_4}$ to be
\begin{equation}
\mathcal{G}_{0,nn11}^{I_1I_2I_3I_4}\propto \delta^{I_1I_2}\delta^{I_3I_4}+ n\sigma U\delta^{I_1I_3}\delta^{I_2I_4}+n\tau UV^{-1}\delta^{I_1I_4}\delta^{I_2I_3}\;.
\end{equation}
For $n=1$, this reduces to (\ref{G0keq1posi}). When $n>1$, we note that GFF only gives the s-channel term proportional to $\delta^{I_1I_2}\delta^{I_3I_4}$. The t and u-channel terms are present because the supergravity fields are dual to a mixture of the single-trace operator and double-trace operators. The mixing is needed to ensure that the extremal three-point functions vanish from the supergravity calculation. 

An important observation is that the above structure based on hidden conformal symmetry only exists for the one-half BPS operators which come from the 6d tensor multiplets. To see this, we can use the same position space method to compute examples of four-point functions where the one-half BPS operators are from the 6d supergravity multiplet. The computation for the one-half BPS operators from the $\Sigma_l$ multiplets is almost identical to that of the $\Theta^I_l$ multiplets. From the selection rules, it is clear that both cases with the same $l\geq 0$ have the same exchange Witten diagrams. Moreover, solving the superconformal Ward identities uniquely determines the contact diagrams in terms of the cubic couplings. However the cubic couplings in these two cases are different \cite{Arutyunov:2000by}. The results for $\Sigma_l$ therefore differ from the tensor four-point functions of $\Theta^I_l$  with $I_i$ set to be equal, and we do not observe a similar structure. 

\section{Future directions}\label{Secfuturedirec}

We conclude with an outline of a few research avenues for the future:
\begin{itemize}

\item The hidden conformal symmetry  enjoyed by four-point tree-level correlators of   tensor  modes in $AdS_3\times S^3$ is in many respects similar to the one that holds in $AdS_5\times S^5$. 
(There is a unique  supermultiplet in 10d IIB  supergravity, and the symmetry holds there for four-point tree-level correlators of {\it all} KK modes).
Both backgrounds are conformally flat, and the relevant flat space superamplitudes enjoy an accidental conformal symmetry, respectively in six and ten dimensions. A third case that should work along very similar lines is $AdS_2 \times S^2$,
where the superamplitude of four external hypermultiplets takes the simple form $G_N^{(4)} \delta^4 (Q) \cdot c$. The kinematic prefactor $G_N^{(4)} \delta^4 (Q)$ is again dimensionless, while $c$ is a just constant, and thus obviously conformally invariant in four dimensions. Assuming the hidden symmetry, it is immediate to write down the generating function of four-point tree-level correlators of all KK modes. It would be interesting to perform some explicit checks of this ansatz. 

\item Another closely related background is $AdS_3\times S^3\times T^4$. Upon reducing IIB supergravity on $T^4$, one obtains $(2,2)$ supergravity in 6d. Our methods can be straightforwardly applied to that case.

\item  It will be important to achieve a first-principles derivation of the hidden conformal symmetry, perhaps along the line of \cite{Maldacena:2011mk}, which related Einstein gravity to conformal gravity.  Such a conceptual 
understanding would also elucidate the regime of validity of the symmetry. For example, does  it extend to higher-point tree-level correlators in $AdS_5 \times S^5$?\footnote{Five-point functions can also be computed using a generalized version of the position space method, see \cite{AdS55pt} for recent progress.} Is it broken by $1/N$ corrections and how?

\item In this paper, we focussed on the four-point functions of the modes $\Theta^I_l$, the KK  tower that arises from the 6d tensor multiplets. Two additional KK towers,  $\Gamma_l$ and $\Sigma_l$,  arise 
from the 6d $(2,0)$ supergraviton. We would like to study the most general four-point functions which involve operators from all these multiplets. We have found that four-point correlators of $\Sigma_l$ fields
are incompatible, at least naively, with 6d conformal symmetry, but perhaps the symmetry is present  in a more subtle way.

\item The full set of  tree-level four-point functions is also needed to solve the mixing problem of double-trace operators and extract the spectrum of anomalous dimensions. These tree-level data can then be used to bootstrap one-loop four-point functions, following the blueprint of \cite{Aharony:2016dwx,Alday:2017xua,Aprile:2017bgs,Aprile:2017xsp,Alday:2017vkk,Aprile:2017qoy}. 
An interesting open question is if the hidden symmetry for the  $\Theta^I_l$ multiplet survives the supergravity loop corrections. 

\item Related to the previous point, it would also be useful to perform an analysis using the Lorentzian inversion formula, along the lines of \cite{Caron-Huot:2018kta}, a method  independent of our supergravity computation.

\end{itemize}

\acknowledgments
We thank Nathan Benjamin, Shai Chester, Wolfger Peelaers, Silviu Pufu and Yifan Wang for discussions. X.Z. also thanks the International Institute of Physics for hospitality during his visit and the participants of the workshop ``Nonperturbative Methods for Conformal Theories'' for useful conversations and comments.  The work of L.R. and K.R. is supported in part by the NSF grant \# PHY1620628.
The work of X.Z. is also supported in part by the Simons Foundation Grant No. 488653.

\appendix 
\section{Twisting small $\mathcal{N}=4$}\label{chiralalgebra}
In this section we give a derivation of the superconformal Ward identities from topological twisting. Let us focus on the global part of the small $\mathcal{N}=4$ superconformal algebra and consider only the holomorphic part. The algebra is $\mathfrak{psu}(1,1|2)$, and is captured by the following commutation relations
\begin{equation}
\begin{split}
{}&[L_0,L_{\pm 1}]=\mp L_{\pm1}\;,\quad [L_1,L_{-1}]=2L_0\;,\quad [J_0^i,J_0^j]=i\epsilon^{ijk}J_0^k\;,\\
{}& [L_0,G^{a A}_{\pm \frac{1}{2}}]=\mp \frac{1}{2}G^{a A}_{\pm \frac{1}{2}}\;,\quad [J_0^i,G^{a A}_{\pm \frac{1}{2}}]=  -\frac{1}{2}(\sigma^i)^a \,_{b}G^{bA}_{\pm \frac{1}{2}}\;,\\
{}& \{G^{a+}_{\frac{1}{2}},G^{b-}_{-\frac{1}{2}})\}=2\epsilon^{ab}L_0-2\sigma_i ^{ab}J^i_0\;,\quad \{G^{a+}_{-\frac{1}{2}},G^{b-}_{\frac{1}{2}}\}=2\epsilon^{ab}L_0+2\sigma_i ^{ab}J^i_0\;,\\
{}& \{G^{a+}_{\frac{1}{2}},G^{b-}_{\frac{1}{2}}\}=2\epsilon^{ab}L_1\;,\quad \{G^{a+}_{-\frac{1}{2}},G^{b-}_{-\frac{1}{2}}\}=2\epsilon^{ab}L_{-1}\;,\\
{}& \{G^{a+}_{\pm\frac{1}{2}},G^{b+}_{\pm\frac{1}{2}}\}=\{G^{a-}_{\pm\frac{1}{2}},G^{b-}_{\pm\frac{1}{2}}\}=0\;,\\
{}&[L_1,G^{ab}_{\frac{1}{2}}]=0\;,\quad[L_1,G^{ab}_{-\frac{1}{2}}]=G^{ab}_{\frac{1}{2}}\;,\quad [L_{-1},G^{ab}_{\frac{1}{2}}]=-G^{ab}_{-\frac{1}{2}}\;,\quad[L_{-1},G^{ab}_{-\frac{1}{2}}]=0\;.
\end{split}
\end{equation}
where $(\sigma_i)^a \,_{b}$ are the Pauli matrices and $\sigma_i^{ab}= (\sigma_i)^a \,_{c}\epsilon^{bc}$ with $\epsilon_{+-}=\epsilon^{+-}=1$. This algebra has an $SU(2)_R$ symmetry as well as an $SU(2)_A$ automorphism under which the supercharges $G_n^{aA}$ transform in $(\bold{2},\bold{2})$.
Following \cite{Beem:2013sza}, we can consider a topological twist by looking at the cohomology of the nilponent supercharge
\begin{equation}
\mathbb{Q}=G^{++}_{-\frac{1}{2}}+G^{-+}_{\frac{1}{2}}\;,\quad \{\mathbb{Q},\mathbb{Q}\}=0\;.
\end{equation}
Operator which are in the $\mathbb{Q}$-cohomology class are one-half BPS under the left-moving $\mathfrak{psu}(1,1|2)$
\begin{equation}
\{\mathbb{Q},\mathcal{O}(0)]=0\;,\quad \mathcal{O}(0)\neq \{\mathbb{Q},O'(0)]\quad \Rightarrow h=j\;.
\end{equation}
Moreover, one can construct the following twisted $\mathfrak{sl}(2,\mathbb{C})$ algebra which is $\mathbb{Q}$-exact
\begin{equation}
\begin{split}
{}&\{\mathbb{Q},G^{--}_{-\frac{1}{2}}\}=2L_{-1}-2\sigma_i^{--}J^i_0\equiv 2\widehat{L}_{-1}\;,\\
{}&\{\mathbb{Q},-G^{+-}_{\frac{1}{2}}\}=2L_1-2\sigma_i^{++} J^i_0\equiv 2\widehat{L}_1\;,\\
{}&\{\mathbb{Q},G^{--}_{\frac{1}{2}}\}=2L_0+2\sigma_i^{+-}J^i_0\equiv 2\widehat{L}_0\;.
\end{split}
\end{equation}
Let us revisit the one-half BPS operators with $SU(2)_L$ indices contracted with spinors
\begin{equation}
\mathcal{O}(z;y)=\mathcal{O}^{\alpha_1,\ldots,\alpha_k}(z)v_{\alpha_1}\ldots v_{\alpha_k}\;, \quad v_\alpha=(1,y)\;.
\end{equation}
When $y=z$, it amounts to inserting operators in nontrivial $\mathbb{Q}$-cohomology classes at the origin and then twist-translating using $\widehat{L}_{-1}$ 
\begin{equation}
\mathcal{O}(z;z)=e^{z\widehat{L}_{-1}}\mathcal{O}(0)e^{-z\widehat{L}_{-1}}\;.
\end{equation}
Because $\widehat{L}_{-1}$  is $\mathbb{Q}$-exact, the twist-translated operators remain in the $\mathbb{Q}$-cohomology. Since the twist construction uses only the left-moving part of the 2d algebra, it commutes with the right-moving algebra. By  standard arguments, the correlators of such twisted operators have no holomorphic dependence. This directly translates into our superconformal Ward identity (\ref{scfwardid}).

\section{Proca-Chern-Simons versus massive Chern-Simons}\label{firstordervectors}
In \cite{Arutyunov:2000by}, it was shown that the vector fields from the spin-2 multiplet $\Gamma_k$ and spin-1 multiplet $\Sigma_k$ satisfy second-order Proca-Chern-Simons equations. Meanwhile their first-order derivatives satisfy a linear constraint and there are only three independent degrees of freedom. In this appendix we show that we can use field redefinition to solve the constraints, which gives three vector fields described by the first-order massive Chern-Simons equations. Moreover, we point out that the magnetic coupling to currents in \cite{Arutyunov:2000by} disappears after the field redefinition. We will work with the cubic vertices where the vector fields couple to two scalar fields $\sigma$. The case with two scalar fields $s^I$ is analogous.

We start from the equation of motion of the gauge fields with quadratic corrections \cite{Arutyunov:2000by}
\begin{equation}\label{fstorderconstraint}
P^\pm_{k-1}A^\pm_\mu+P^\pm_{k+3}C^\pm_\mu=\pm(W^{\sigma\sigma A^\pm}_{123}+W^{\sigma\sigma C^\pm}_{123})J_\mu\;,
\end{equation}
\begin{equation}\label{sndordereom}
P^\mp_{k+1}P^\pm_{k-1}A^\pm_\mu-P^\mp_{k+1}P^\pm_{k+3}C^\pm_\mu=(V^{\sigma\sigma A^\pm}_{123}-V^{\sigma\sigma C^\pm}_{123})J_\mu\pm(W^{\sigma\sigma A^\pm}_{123}P^\pm_{k-1}J_\mu-W^{\sigma\sigma C^\pm}_{123}P^\pm_{k+3}J_\mu)\; ,
\end{equation}
where 
\begin{equation}
J_\mu=\partial_\mu \sigma_1\sigma_2- \sigma_1\partial_\mu\sigma_2\;,
\end{equation}
and $P^\pm_m$ is the differential operator 
\begin{equation}
(P_m^{\pm})_\mu{}^\lambda=\epsilon_\mu{}^{\nu\lambda}\triangledown_\nu\pm m\delta^\lambda_\mu\;.
\end{equation}
The coefficients $W^{\sigma\sigma A^\pm}_{123}$, $W^{\sigma\sigma C^\pm}_{123}$, $V^{\sigma\sigma A^\pm}_{123}$, $V^{\sigma\sigma C^\pm}_{123}$ are defined in \cite{Arutyunov:2000by} but their precisely forms are not important to us. One can act with $P^\mp_{k+1}$ on the first equation and solve for one variable in the second equation to get Proca-Chern-Simons equations for $A^\pm_\mu$ and $C^\pm_\mu$. We also notice that the couplings to the current $J_\mu$ are both electric and magnetic ({\it i.e.}, $V^\mu J_\mu$ and $\epsilon^{\mu\nu\rho}V_\mu\triangledown_\nu J_\rho$).

We now define the following new field 
\begin{equation}\label{defL}
L^\pm_\mu=\pm\frac{1}{2}(P_0A_\mu-P_0C_\mu)+\frac{2 (k_1-1) (k_2-1) (k+1)}{(k_1+1) (k_2+1)}J_\mu\;.
\end{equation}
In terms of $L^\pm_\mu$ we can rewrite (\ref{sndordereom}) as
\begin{equation}\label{sndtofstordereom}
\pm 2P_0L^\pm_\mu\mp2A^\pm_\mu\mp2C^\pm_\mu-(k-1)(k+1)A^\pm_\mu+(k+1)(k+3)C^\pm_\mu=\mathcal{Q}^\pm_\mu(
\sigma)\; ,
\end{equation}
where the $\mathcal{Q}^\pm_\mu(
\sigma)$ on the RHS is the quadratic corrections from $\sigma$ and does not depend on $A^\pm_\mu$, $C^\pm_\mu$, $L^\pm_\mu$. The explicit expression of $\mathcal{Q}^\pm_\mu(
\sigma)$ can be derived from above, but we will not write it down here. Therefore, in terms of $A^\pm_\mu$, $C^\pm_\mu$, $L^\pm_\mu$ we have a system of first-order massive Chern-Simons equations (\ref{defL}), (\ref{fstorderconstraint}), (\ref{sndtofstordereom}) which can be more compactly written as 
\begin{equation}
(P_0\mathbf{1}+\mathbf{M}^\pm)\left(\begin{array}{c}L^\pm_\mu \\A^\pm_\mu \\C^\pm_\mu\end{array}\right)=\mathbf{Q}^\pm_\mu(\sigma)\;.
\end{equation}
Here $\mathbf{1}$ is the unit matrix and $\mathbf{M}^\pm$ is a mass matrix. The vector $\mathbf{Q}^\pm_\mu(\sigma)$ contains all the other terms depending on $\sigma$. The mass matrix can be diagonalized by using the following linear combinations
\begin{eqnarray}
V^\pm_{1,\mu}&=&\frac{k(1-k)L^\pm_\mu-2(k+1)^2A^\pm_\mu+k(k+3)C^\pm_\mu}{2^{4/3}k (k+2)^{1/3}}\;,\\
V^\pm_{2,\mu}&=& \frac{k L^\pm_\mu-(k+2)A^\pm_\mu}{2^{4/3}k (k+2)^{1/3}}\;,\\
V^\pm_{3,\mu}&=& \frac{A^\pm_\mu+C^\pm_\mu}{2^{4/3}k (k+2)^{1/3}}\;.
\end{eqnarray}
In terms of $V^\pm_{i,\mu}$, we have 
\begin{equation}
(P_0\mathbf{1}\pm\mathbf{\Lambda})\left(\begin{array}{c}V^\pm_{1,\mu} \\V^\pm_{2,\mu} \\V^\pm_{3,\mu}\end{array}\right)=\mathbf{Q}'^\pm_\mu(\sigma)\;.
\end{equation}
where $\mathbf{\Lambda}={\rm diag}\{k-1,-k-1,k+3\}$ is the diagonalized mass matrix. Moreover, as a result of (\ref{defL}) one can check that there is only only electric coupling  in $\mathbf{Q}'^\pm_\mu(\sigma)$, {\it i.e.}, no $\epsilon^{\mu\nu\rho}\triangledown_\nu J_\rho$ appears. These eigenvectors are the fields appeared in our tables \ref{tablegravtion}, \ref{tablescalarG}. We can identify $V^\pm_{1,\mu}$, $V^\pm_{2,\mu}$, $V^\pm_{3,\mu}$ respectively with $V^\pm_\mu$ (at level $k-1$), $Y^\mp_\mu$ (at level $k-1$), $W^\pm_\mu$ (at level $k+1$).

\section{$AdS_3$ Witten diagrams }\label{AdS3Wdiagrams}
In this appendix we discuss the computation of exchange Witten diagrams which are unique to $AdS_3$. Exchange diagrams of other fields, such as scalars, Proca fields, massive symmetric traceless tensors can be computed using the standard method. See, {\it e.g.}, Appendix A of \cite{Rastelli:2017udc} for a summary of formulae.

\subsection{Contact Witten diagrams with three derivatives}
Before we start discussing exchange Witten diagrams, it is useful to first look at a special type of contact Witten diagrams which has an odd number of derivatives. The special contact Witten diagram is built from contact vertices of the type 
\begin{equation}
\epsilon^{\mu\nu\rho}\left(\partial_\mu\phi_1\partial_\nu\phi_2\partial_\rho\phi_3\right) \phi_4\phi_5\ldots\;,
\end{equation}
and has {\it three} derivatives. We will focus on the case where only four scalar fields are involved, though it is straightforward to generalize the result to include more scalar fields. We are looking at the following contact Witten diagram (we have distributed the three derivatives on the external legs 1, 3 and 4) defined by the integral 
\begin{equation}
W_{\rm con}^{(134)}\equiv \int \frac{d^3z}{z_0^3}z_0^3 \epsilon_{\mu\nu\rho} \partial_\mu G_{B\partial}^{\Delta_1}(z,x_1)\partial_\nu G_{B\partial}^{\Delta_3}(z,x_3)\partial_\rho G_{B\partial}^{\Delta_4}(z,x_4)G_{B\partial}^{\Delta_2}(z,x_2)\; ,
\end{equation}
where $G_{B\partial}^{\Delta_i}(z,x_i)$ is the bulk-to-boundary propagator 
\begin{equation}
G_{B\partial}^{\Delta_i}(z,x_i)=\left(\frac{z_0}{z_0^2+(\vec{z}-\vec{x_i})^2}\right)^{\Delta_i}\;.
\end{equation}
This integral can be evaluated to the following 
\begin{equation}
\begin{split}
{}&W_{\rm con}^{(134)}=x_{13}^{\Delta_4-\Delta_1-\Delta_2-\Delta_3}x_{14}^{-\Delta_1+\Delta_2+\Delta_3-\Delta_4}x_{24}^{2\Delta_2}r_{34}^{\Delta_1+\Delta_2-\Delta_3-\Delta_4}\frac{i \pi \Gamma(\frac{\Delta_1+\Delta_2+\Delta_3+\Delta_4}{2})}{\Gamma(\Delta_1)\Gamma(\Delta_2)\Gamma(\Delta_3)\Gamma(\Delta_4)}\\ 
{}&\quad\quad\times (z-\bar{z})(\bar{D}_{\Delta_1+2,\Delta_2,\Delta_3+1,\Delta_4+1}+\bar{D}_{\Delta_1+1,\Delta_2,\Delta_3+2,\Delta_4+1}-\Delta_4\bar{D}_{\Delta_1+1,\Delta_2,\Delta_3+1,\Delta_4})\;,
\end{split}
\end{equation}
where $z$ and $\bar{z}$ are the chiral and anti-chiral cross ratios. The $\bar{D}$-functions are related to the standard $D$-functions 
\begin{equation}
D_{\Delta_1\Delta_2\Delta_3\Delta_4}\equiv \int \frac{d^3z}{z_0^3} G_{B\partial}^{\Delta_1}(z,x_1)G_{B\partial}^{\Delta_2}(z,x_2)G_{B\partial}^{\Delta_3}(z,x_3) G_{B\partial}^{\Delta_4}(z,x_4)\;,
\end{equation}
via 
\begin{equation}\label{dbar}
\frac{ \prod_{i=1}^4\Gamma(\Delta_i)}{\Gamma(\Sigma-\frac{1}{2}d)}\frac{2}{\pi^{\frac{d}{2}}}D_{\Delta_1\Delta_2\Delta_3\Delta_4}(x_1,x_2,x_3,x_4)=\frac{r_{14}^{\Sigma-\Delta_1-\Delta_4}r_{34}^{\Sigma-\Delta_3-\Delta_4}}{r_{13}^{\Sigma-\Delta_4}r_{24}^{\Delta_2}}\bar{D}_{\Delta_1\Delta_2\Delta_3\Delta_4} (U,V),
\end{equation}
where $d=2$ and $2\Sigma \equiv \sum_{i=1}^4 \Delta_i$.

\subsection{Exchange Witten diagrams of massless and massive Chern-Simons fields}
We now move on to the exchange Witten diagram of massless and massive Chern-Simons fields. The propagator satisfies the equation of motion
\begin{equation}\label{mCSpropgator}
P_{k-1}^\pm G^{mCS,k,\pm}_{\mu;\nu}(z_1,z_2)=\mp g_{\mu\nu}\delta(z_1,z_2)+(\ldots)\delta_{k,1} \; ,
\end{equation}
where $\ldots$ are suitable terms added to make the differential operator invertible \cite{DHoker:1999bve}. The corresponding vector field has conformal dimension $\Delta=k$. When $k=1$, the vector field is a massless gauge field in the bulk and when $k\neq 1$ the vector is massive. The exchange diagram is defined by
\begin{equation}
W^{CS,k,\pm}=\int \frac{d^3z}{z_0^3}\frac{d^3w}{w_0^3}J^\mu(z;x_1,x_2)G^{mCS,k,\pm}_{\mu;\nu}(z,w)J^\nu(w;x_3,x_4) \; ,
\end{equation}
where $J_\mu(z;x_1,x_2)$ is a conserved current made out of scalar bulk-to-boundary propagators
\begin{equation}
J_\mu(z;x_1,x_2)=\partial_\mu G^{\Delta_\phi}_{B\partial}(z,x_1)G^{\Delta_\phi}_{B\partial}(z,x_2)-G^{\Delta_\phi}_{B\partial}(z,x_1)\partial_\mu G^{\Delta_\phi}_{B\partial}(z,x_2)\;,\quad \triangledown^\mu J_\mu=0\;.
\end{equation}
The massive case can be treated using the method of \cite{DHoker:1999mqo} with slight modifications. The massless case however requires special attention, and the method of \cite{DHoker:1999mqo} leads to formal divergences. In this subsection, we will give a different method to compute these exchange diagrams which can be applied to both the massless and the massive case. For simplicity, we will restrict ourselves to the case where the external operators have the same dimension $\Delta_i=\Delta_\phi$.

The idea is to view the exchange Witten diagram as solution to a differential equation with certain boundary conditions. We first look at Witten exchange diagrams of Proca fields as a more familiar example. The exchange diagrams are defined by 
\begin{equation}\label{Wproca}
W_{{\rm Proca},k}=\int \frac{d^3z}{z_0^3}\frac{d^3w}{w_0^3}J^\mu(z;x_1,x_2)G_{{\rm Proca}, k,\mu;\nu}(z,w)J^\nu(w;x_3,x_4) \; ,
\end{equation}
The propagator $G_{{\rm Proca},k}$ of a Proca field with squared-mass $m_k^2=(k-1)^2-2$ (and dual dimension $\Delta_k=k$) satisfies
\begin{equation}
\left({\rm Proca}_{k} G_{{\rm Proca}, k}\right)_{\mu;\nu}=g_{\mu\nu}\delta(z_1,z_2)+(\ldots)\delta_{k,1} \; ,
\end{equation}
where 
\begin{equation}
({\rm Proca}_{k})_\mu{}^\nu\equiv \square\delta_\mu^\nu-g^{\nu\rho}\triangledown_\rho\triangledown_\mu-m_k^2 \delta_\mu^\nu\;.
\end{equation}
Now consider only the $z$ integral in (\ref{Wproca}) 
\begin{equation}
I_{{\rm Proca},k,\nu}(x_1,x_2;w)\equiv\int \frac{d^3z}{z_0^3}J^\mu(z;x_1,x_2)G_{{\rm Proca}, k,\mu;\nu}(z,w)\;.
\end{equation}
We act on the integral with the differential operator 
\begin{equation}
\mathbf{L}^{(1)}_{AB}+\mathbf{L}^{(2)}_{AB}+\mathcal{L}^{(w)}_{AB} \; ,
\end{equation}
where $\mathbf{L}^{(i)}_{AB}$ are the conformal generators of the boundary point $x_i$, and $\mathcal{L}^{(w)}_{AB}$ is AdS isometry generator of the bulk point $w$. Because the $z$-integral is conformally covariant, it is annihilated by this operator
\begin{equation}
(\mathbf{L}^{(1)}_{AB}+\mathbf{L}^{(2)}_{AB}+\mathcal{L}^{(w)}_{AB})I_{{\rm Proca},k,\nu}(x_1,x_2;w)=0\;.
\end{equation}
It follows that 
\begin{equation}
-\frac{1}{2}\left(\mathbf{L}^{(1)}_{AB}+\mathbf{L}^{(2)}_{AB}\right)^2\delta_\mu^\nu I_{{\rm Proca},k,\nu}(x_1,x_2;w)=-\frac{1}{2}\left(\mathcal{L}^{(w)}_{AB}\right)^2\delta_\mu^\nu I_{{\rm Proca},k,\nu}(x_1,x_2;w)\;.
\end{equation}
Note that the operator on the LHS is nothing but the two-particle quadratic Casimir. The operator on the RHS is the AdS Laplacian with a constant shift \cite{Pilch:1984xx}
\begin{equation}
-\frac{1}{2}\left(\mathcal{L}^{(z)}_{AB}\right)^2\delta_\mu^\nu=(\square+2)\delta_\mu^\nu=({\rm Proca}_{k})_\mu{}^\nu+g^{\nu\rho}\triangledown_\rho\triangledown_\mu+(k-1)^2\delta_\mu^\nu\;.
\end{equation}
Now we can use the equation of motion of the bulk-to-bulk propagator and perform the remaining $w$-integral. The operator $g^{\nu\rho}\triangledown_\rho\triangledown_\mu$ can be ignored because we can integrate by part. Its contribution vanishes since $I_{{\rm Proca},k,\nu}(x_1,x_2;w)$ is coupled to a conserved current. All in all, we get 
\begin{equation}\label{CasimirProca}
\left(\mathbf{Casimir}^{(12)}-(k-1)^2 \right)W_{{\rm Proca},k}=W^{\rm con}_{\rm Proca} \; ,
\end{equation}
where $W^{\rm con}_{\rm Proca}$ is a two-derivative contact diagram 
\begin{equation}
W^{\rm con}_{\rm Proca}=\int \frac{d^3z}{z_0^3}J^\mu(z;x_1,x_2)J_\mu(z;x_3,x_4)\;.
\end{equation}
Instead of evaluating the diagram using the method of \cite{DHoker:1999mqo}, we can alternatively solve the differential equation (\ref{CasimirProca}). We first need a special solution. This is not difficult for the cases when the method of \cite{DHoker:1999mqo} applies, and the answer is a finite sum of D-functions. The equation (\ref{CasimirProca}) has two homogenous solutions, which are the conformal block of the exchanged single-trace operator and its shadow. They can be fixed by imposing boundary conditions. When we decompose $W^{\rm con}_{\rm Proca}$, it should contain only single-trace and double-trace blocks, and no shadow conformal block. Moreover in the Euclidean regime, {\it i.e.}, $\bar{z}=z^*$, $W^{\rm con}_{\rm Proca}$ is single-valued (as is clear from its integral definition). These conditions uniquely fix the solution.

To compute the massive Chern-Simons exchange diagrams, we first notice the following relations among differential operators  
\begin{equation}
(P^-_{k-1} P^+_{k-1})_\mu{}^\nu=\underbrace{\square\delta_\mu^\nu-\triangledown_\mu \triangledown^\nu-((k-1)^2-2) \delta_\mu^\nu}_{({\rm Proca}_k)_\mu{}^\nu}\;.
\end{equation}
It then follows from (\ref{mCSpropgator}) that the massive Chern-Simons propagator can be obtained from applying $P^\mp_{k-1}$ on  (Maxwell) Proca propagators
\begin{equation}\label{mCSandProca}
G^{{\rm mCS},\pm,k}_{\mu;\nu}=\mp (P_{k-1}^{\mp} G_{{\rm Proca},k})_{\mu;\nu}\;.
\end{equation}
We now act on the massive Chern-Simons exchange diagram with the two-particle quadratic Casimir. Using the same argument and  (\ref{mCSandProca}), we get the following differential equation
\begin{equation}\label{CasimirmCS}
\left(\mathbf{Casimir}^{(12)}-(k-1)^2 \right)W^{CS,k,\pm}=W^{\rm con}_{{\rm mCS},k}\; ,
\end{equation}
where 
\begin{equation}
W^{\rm con}_{{\rm mCS},k}=\int \frac{d^3z}{z_0^3}J^\mu(z;x_1,x_2)(P^\mp_{k-1}J)_\mu(z;x_3,x_4)=\pm 2(W_{\rm con}^{(134)}-W_{\rm con}^{(234)})+(k-1)W^{\rm con}_{\rm Proca}\;.
\end{equation}
Note that for $k=1$, {\it i.e.}, the massless case, there are only three-derivative contact terms. When $k>1$, we can write 
\begin{equation}
W^{\rm con}_{{\rm mCS},k}=\widetilde{W}^{\rm con}_{{\rm mCS},k}+ \frac{1}{k-1}W_{{\rm Proca},k}\;,
\end{equation}
so that the differential equation for $\widetilde{W}^{\rm con}_{{\rm mCS},k}$ reduces to the massless form. The special solutions are again easy to guess, and take the general form of $(z-\bar{z})$ times a sum of $D$-functions. We will list a few explicit solutions in a moment. The equation (\ref{CasimirmCS}) also admit homogenous solutions which are the conformal block for the single-trace operator and its shadow operator. To fix the solution we require that in the conformal block decomposition, the single-trace conformal block has dimension $(h,\bar{h})=\left(\frac{k+1}{2},\frac{k-1}{2}\right)$ for $+$, and $(h,\bar{h})=\left(\frac{k-1}{2},\frac{k+1}{2}\right)$ for $-$. There is no shadow conformal block. The solution is also single-valued in the Euclidean regime. 

Using this method, we can easily compute the massless and massive Chern-Simons exchange diagrams. Let us list the values of the diagrams which appear in this paper.\footnote{We have rescaled the exchange diagrams by some overall factors which are unimportant to the position space method.} 
\subsubsection*{\underline{$\Delta_\phi=1$}}
\begin{equation}
W^{CS,1,\pm}=\frac{1}{x_{12}^2 x_{34}^2}\frac{\pi}{2}\left(\mp (z-\bar{z})U\bar{D}_{2211}+\log V\right)\;.
\end{equation}

\subsubsection*{\underline{$\Delta_\phi=2$}}
\begin{equation}
W^{CS,1,\pm}=\frac{\pi}{x_{12}^4 x_{34}^4}\left(\mp (z-\bar{z})U^2(\bar{D}_{3311}+2\bar{D}_{3322})+\log V\right)\;,
\end{equation}

\begin{equation}
W^{CS,3,\pm}=\frac{\pi U}{x_{12}^4 x_{34}^4}\left(\pm 2(z-\bar{z})U\bar{D}_{3322}+\bar{D}_{1223}-V\bar{D}_{1232}-\bar{D}_{2123}+\bar{D}_{2132}\right)\;.
\end{equation}

\subsubsection*{\underline{$\Delta_\phi=3$}}
\begin{equation}
W^{CS,1,\pm}=\frac{\pi}{x_{12}^6 x_{34}^6}\left(\mp (z-\bar{z})U^3(\bar{D}_{4411}+3\bar{D}_{4422}+3\bar{D}_{4433})+2\log V\right)\;,
\end{equation}

\begin{equation}
\begin{split}
W^{CS,3,\pm}={}&\frac{\pi U}{x_{12}^6 x_{34}^6}\bigg(\pm2(z-\bar{z})U^2(3\bar{D}_{4422}+4\bar{D}_{4433})+3(\bar{D}_{1234}-V\bar{D}_{1243}-\bar{D}_{2134}+\bar{D}_{2143})\\
{}&+4U(\bar{D}_{2334}-V\bar{D}_{2343}-\bar{D}_{3234}+\bar{D}_{3243})\bigg)\;,
\end{split}
\end{equation}

\begin{equation}
W^{CS,5,\pm}=\frac{\pi U^2}{x_{12}^6 x_{34}^6}\left(\pm(z-\bar{z})U\bar{D}_{4433}+\bar{D}_{2334}-V\bar{D}_{2343}-\bar{D}_{3234}+\bar{D}_{3243}\right)\;.
\end{equation}

\subsubsection*{\underline{$\Delta_\phi=4$}}
\begin{equation}
W^{CS,1,\pm}=\frac{\pi}{x_{12}^8 x_{34}^8}\left(\mp (z-\bar{z})U^4(3\bar{D}_{5511}+12\bar{D}_{5522}+15\bar{D}_{5533}+10\bar{D}_{5544})+18\log V\right)\;,
\end{equation}

\begin{equation}
\begin{split}
W^{CS,3,\pm}={}&\frac{\pi U}{x_{12}^8 x_{34}^8}\bigg(\pm2(z-\bar{z})U^3(12\bar{D}_{5522}+20\bar{D}_{5533}+15\bar{D}_{5544})\\
+{}&12(\bar{D}_{1245}-V\bar{D}_{1254}-\bar{D}_{2145}+\bar{D}_{2154})+20U(\bar{D}_{2345}-V\bar{D}_{2354}-\bar{D}_{3245}+\bar{D}_{3254})	\\
+{}&15U^2(\bar{D}_{3445}-V\bar{D}_{3454}-\bar{D}_{4345}+\bar{D}_{4354})\bigg)\;,
\end{split}
\end{equation}

\begin{equation}
\begin{split}
W^{CS,5,\pm}={}&\frac{\pi U^2}{x_{12}^8 x_{34}^8}\bigg(\pm (z-\bar{z})U^2(5\bar{D}_{5533}+6\bar{D}_{5544})+5(
\bar{D}_{2345}-V\bar{D}_{2354}-\bar{D}_{3245}+\bar{D}_{3254})\\
+{}&6U(\bar{D}_{3445}-V\bar{D}_{3454}-\bar{D}_{4345}+\bar{D}_{4354})\bigg)\;,
\end{split}
\end{equation}

\begin{equation}
W^{CS,7,\pm}=\frac{\pi U^3}{x_{12}^8 x_{34}^8}\left(\pm 2(z-\bar{z})U\bar{D}_{5544}+3(\bar{D}_{3445}-V\bar{D}_{3454}-\bar{D}_{4345}+\bar{D}_{4354})\right)\;.
\end{equation}

\subsection{Exchange Witten diagrams of non-dynamical graviton field}
The exchange Witten diagrams of graviton field in $AdS_3$ also cannot be evaluated using the method of \cite{DHoker:1999mqo}. However, it is straightforward to adapt the method from the previous subsection to the case of non-dynamical gravitons. We will not repeat the analysis but simply write down the solutions for reader's reference.

\subsubsection*{\underline{$\Delta_\phi=1$}}
\begin{equation}
W_{gr}=\frac{\pi}{2x_{12}^2x_{34}^2}(2+U(U-V-1)\bar{D}_{2211})\;.
\end{equation} 

\subsubsection*{\underline{$\Delta_\phi=2$}}
\begin{equation}
W_{gr}=\frac{\pi}{x_{12}^4x_{34}^4}\bigg(4+U^2\big(-\bar{D}_{2211}-5\bar{D}_{2222}+(U-V-1)(2\bar{D}_{3311}+3\bar{D}_{3322})\big)\bigg)\;.
\end{equation} 

\subsubsection*{\underline{$\Delta_\phi=3$}}
\begin{equation}
\begin{split}
W_{gr}={}&\frac{\pi}{x_{12}^6x_{34}^6}\bigg(12+U^3\big(-3\bar{D}_{3311}-4\bar{D}_{3322}-16\bar{D}_{3333}\\
+{}&(U-V-1)(3\bar{D}_{4411}+6\bar{D}_{4422}+5\bar{D}_{4433})\big)\bigg)\;.
\end{split}
\end{equation} 

\subsubsection*{\underline{$\Delta_\phi=4$}}
\begin{equation}
\begin{split}
W_{gr}={}&\frac{\pi}{x_{12}^8x_{34}^8}\bigg(288+U^4\big(-36\bar{D}_{4411}-60\bar{D}_{4422}-45\bar{D}_{4433}-165\bar{D}_{4444}\\
+{}&(U-V-1)(24\bar{D}_{5511}+60\bar{D}_{5522}+60\bar{D}_{5533}+35\bar{D}_{5544})\big)\bigg)\;.
\end{split}
\end{equation}

\bibliography{AdS3fourpoint} 

\providecommand{\href}[2]{#2}\begingroup\raggedright\begin{thebibliography}{10}

\bibitem{Rastelli:2016nze}
L.~Rastelli and X.~Zhou, ``{Mellin amplitudes for $AdS_5\times S^5$},''
  \href{http://dx.doi.org/10.1103/PhysRevLett.118.091602}{{\em Phys. Rev.
  Lett.} {\bfseries 118} no.~9, (2017) 091602},
\href{http://arxiv.org/abs/1608.06624}{{\ttfamily arXiv:1608.06624 [hep-th]}}.

\bibitem{Rastelli:2017udc}
L.~Rastelli and X.~Zhou, ``{How to Succeed at Holographic Correlators Without
  Really Trying},'' \href{http://dx.doi.org/10.1007/JHEP04(2018)014}{{\em JHEP}
  {\bfseries 04} (2018) 014},
\href{http://arxiv.org/abs/1710.05923}{{\ttfamily arXiv:1710.05923 [hep-th]}}.

\bibitem{Arutyunov:2017dti}
G.~Arutyunov, S.~Frolov, R.~Klabbers, and S.~Savin, ``{Towards 4-point
  correlation functions of any $ \frac{1}{2} $ -BPS operators from
  supergravity},'' \href{http://dx.doi.org/10.1007/JHEP04(2017)005}{{\em JHEP}
  {\bfseries 04} (2017) 005},
\href{http://arxiv.org/abs/1701.00998}{{\ttfamily arXiv:1701.00998 [hep-th]}}.

\bibitem{Arutyunov:2018tvn}
G.~Arutyunov, R.~Klabbers, and S.~Savin, ``{Four-point functions of 1/2-BPS
  operators of any weights in the supergravity approximation},''
  \href{http://dx.doi.org/10.1007/JHEP09(2018)118}{{\em JHEP} {\bfseries 09}
  (2018) 118},
\href{http://arxiv.org/abs/1808.06788}{{\ttfamily arXiv:1808.06788 [hep-th]}}.

\bibitem{Arutyunov:2018neq}
G.~Arutyunov, R.~Klabbers, and S.~Savin, ``{Four-point functions of
  all-different-weight chiral primary operators in the supergravity
  approximation},'' \href{http://dx.doi.org/10.1007/JHEP09(2018)023}{{\em JHEP}
  {\bfseries 09} (2018) 023},
\href{http://arxiv.org/abs/1806.09200}{{\ttfamily arXiv:1806.09200 [hep-th]}}.

\bibitem{Alday:2017xua}
L.~F. Alday and A.~Bissi, ``{Loop Corrections to Supergravity on $AdS_5 \times
  S^5$},'' \href{http://dx.doi.org/10.1103/PhysRevLett.119.171601}{{\em Phys.
  Rev. Lett.} {\bfseries 119} no.~17, (2017) 171601},
\href{http://arxiv.org/abs/1706.02388}{{\ttfamily arXiv:1706.02388 [hep-th]}}.

\bibitem{Aprile:2017bgs}
F.~Aprile, J.~M. Drummond, P.~Heslop, and H.~Paul, ``{Quantum Gravity from
  Conformal Field Theory},''
  \href{http://dx.doi.org/10.1007/JHEP01(2018)035}{{\em JHEP} {\bfseries 01}
  (2018) 035},
\href{http://arxiv.org/abs/1706.02822}{{\ttfamily arXiv:1706.02822 [hep-th]}}.

\bibitem{Aprile:2017xsp}
F.~Aprile, J.~M. Drummond, P.~Heslop, and H.~Paul, ``{Unmixing Supergravity},''
  \href{http://dx.doi.org/10.1007/JHEP02(2018)133}{{\em JHEP} {\bfseries 02}
  (2018) 133},
\href{http://arxiv.org/abs/1706.08456}{{\ttfamily arXiv:1706.08456 [hep-th]}}.

\bibitem{Aprile:2018efk}
F.~Aprile, J.~Drummond, P.~Heslop, and H.~Paul, ``{Double-trace spectrum of
  $N=4$ supersymmetric Yang-Mills theory at strong coupling},''
  \href{http://dx.doi.org/10.1103/PhysRevD.98.126008}{{\em Phys. Rev.}
  {\bfseries D98} no.~12, (2018) 126008},
\href{http://arxiv.org/abs/1802.06889}{{\ttfamily arXiv:1802.06889 [hep-th]}}.

\bibitem{Caron-Huot:2018kta}
S.~Caron-Huot and A.-K. Trinh, ``{All tree-level correlators in
  AdS$_{5}$×S$_{5}$ supergravity: hidden ten-dimensional conformal
  symmetry},'' \href{http://dx.doi.org/10.1007/JHEP01(2019)196}{{\em JHEP}
  {\bfseries 01} (2019) 196},
\href{http://arxiv.org/abs/1809.09173}{{\ttfamily arXiv:1809.09173 [hep-th]}}.

\bibitem{Aharony:2016dwx}
O.~Aharony, L.~F. Alday, A.~Bissi, and E.~Perlmutter, ``{Loops in AdS from
  Conformal Field Theory},''
  \href{http://dx.doi.org/10.1007/JHEP07(2017)036}{{\em JHEP} {\bfseries 07}
  (2017) 036},
\href{http://arxiv.org/abs/1612.03891}{{\ttfamily arXiv:1612.03891 [hep-th]}}.

\bibitem{Aprile:2017qoy}
F.~Aprile, J.~M. Drummond, P.~Heslop, and H.~Paul, ``{Loop corrections for
  Kaluza-Klein AdS amplitudes},''
  \href{http://dx.doi.org/10.1007/JHEP05(2018)056}{{\em JHEP} {\bfseries 05}
  (2018) 056},
\href{http://arxiv.org/abs/1711.03903}{{\ttfamily arXiv:1711.03903 [hep-th]}}.

\bibitem{Alday:2017vkk}
L.~F. Alday and S.~Caron-Huot, ``{Gravitational S-matrix from CFT dispersion
  relations},'' \href{http://dx.doi.org/10.1007/JHEP12(2018)017}{{\em JHEP}
  {\bfseries 12} (2018) 017},
\href{http://arxiv.org/abs/1711.02031}{{\ttfamily arXiv:1711.02031 [hep-th]}}.

\bibitem{Goncalves:2014ffa}
V.~Gon{\c c}alves, ``{Four point function of $\mathcal{N}=4$ stress-tensor
  multiplet at strong coupling},''
  \href{http://dx.doi.org/10.1007/JHEP04(2015)150}{{\em JHEP} {\bfseries 04}
  (2015) 150},
\href{http://arxiv.org/abs/1411.1675}{{\ttfamily arXiv:1411.1675 [hep-th]}}.

\bibitem{Alday:2018pdi}
L.~F. Alday, A.~Bissi, and E.~Perlmutter, ``{Genus-One String Amplitudes from
  Conformal Field Theory},''
\href{http://arxiv.org/abs/1809.10670}{{\ttfamily arXiv:1809.10670 [hep-th]}}.

\bibitem{Alday:2018kkw}
L.~F. Alday, ``{On Genus-one String Amplitudes on $AdS_5 \times S^5$},''
\href{http://arxiv.org/abs/1812.11783}{{\ttfamily arXiv:1812.11783 [hep-th]}}.

\bibitem{Binder:2019jwn}
D.~J. Binder, S.~M. Chester, S.~S. Pufu, and Y.~Wang, ``{$\mathcal{N}=4$
  Super-Yang-Mills Correlators at Strong Coupling from String Theory and
  Localization},''
\href{http://arxiv.org/abs/1902.06263}{{\ttfamily arXiv:1902.06263 [hep-th]}}.

\bibitem{Rastelli:2017ymc}
L.~Rastelli and X.~Zhou, ``{Holographic Four-Point Functions in the (2, 0)
  Theory},'' \href{http://dx.doi.org/10.1007/JHEP06(2018)087}{{\em JHEP}
  {\bfseries 06} (2018) 087},
\href{http://arxiv.org/abs/1712.02788}{{\ttfamily arXiv:1712.02788 [hep-th]}}.

\bibitem{Zhou:2017zaw}
X.~Zhou, ``{On Superconformal Four-Point Mellin Amplitudes in Dimension
  $d>2$},'' \href{http://dx.doi.org/10.1007/JHEP08(2018)187}{{\em JHEP}
  {\bfseries 08} (2018) 187},
\href{http://arxiv.org/abs/1712.02800}{{\ttfamily arXiv:1712.02800 [hep-th]}}.

\bibitem{Heslop:2017sco}
P.~Heslop and A.~E. Lipstein, ``{M-theory Beyond The Supergravity
  Approximation},'' \href{http://dx.doi.org/10.1007/JHEP02(2018)004}{{\em JHEP}
  {\bfseries 02} (2018) 004},
\href{http://arxiv.org/abs/1712.08570}{{\ttfamily arXiv:1712.08570 [hep-th]}}.

\bibitem{Zhou:2018ofp}
X.~Zhou, ``{On Mellin Amplitudes in SCFTs with Eight Supercharges},''
  \href{http://dx.doi.org/10.1007/JHEP07(2018)147}{{\em JHEP} {\bfseries 07}
  (2018) 147},
\href{http://arxiv.org/abs/1804.02397}{{\ttfamily arXiv:1804.02397 [hep-th]}}.

\bibitem{Chester:2018dga}
S.~M. Chester and E.~Perlmutter, ``{M-Theory Reconstruction from (2,0) CFT and
  the Chiral Algebra Conjecture},''
  \href{http://dx.doi.org/10.1007/JHEP08(2018)116}{{\em JHEP} {\bfseries 08}
  (2018) 116},
\href{http://arxiv.org/abs/1805.00892}{{\ttfamily arXiv:1805.00892 [hep-th]}}.

\bibitem{Abl:2019jhh}
T.~Abl, P.~Heslop, and A.~E. Lipstein, ``{Recursion relations for anomalous
  dimensions in the 6d $(2, 0)$ theory},''
  \href{http://dx.doi.org/10.1007/JHEP04(2019)038}{{\em JHEP} {\bfseries 04}
  (2019) 038},
\href{http://arxiv.org/abs/1902.00463}{{\ttfamily arXiv:1902.00463 [hep-th]}}.

\bibitem{Chester:2018lbz}
S.~M. Chester, ``{AdS$_{4}$/CFT$_{3}$ for unprotected operators},''
  \href{http://dx.doi.org/10.1007/JHEP07(2018)030}{{\em JHEP} {\bfseries 07}
  (2018) 030},
\href{http://arxiv.org/abs/1803.01379}{{\ttfamily arXiv:1803.01379 [hep-th]}}.

\bibitem{Chester:2018aca}
S.~M. Chester, S.~S. Pufu, and X.~Yin, ``{The M-Theory S-Matrix From ABJM:
  Beyond 11D Supergravity},''
  \href{http://dx.doi.org/10.1007/JHEP08(2018)115}{{\em JHEP} {\bfseries 08}
  (2018) 115},
\href{http://arxiv.org/abs/1804.00949}{{\ttfamily arXiv:1804.00949 [hep-th]}}.

\bibitem{Binder:2018yvd}
D.~J. Binder, S.~M. Chester, and S.~S. Pufu, ``{Absence of $D^4 R^4$ in
  M-Theory From ABJM},''
\href{http://arxiv.org/abs/1808.10554}{{\ttfamily arXiv:1808.10554 [hep-th]}}.

\bibitem{Mack:2009mi}
G.~Mack, ``{D-independent representation of Conformal Field Theories in D
  dimensions via transformation to auxiliary Dual Resonance Models. Scalar
  amplitudes},''
\href{http://arxiv.org/abs/0907.2407}{{\ttfamily arXiv:0907.2407 [hep-th]}}.

\bibitem{Penedones:2010ue}
J.~Penedones, ``{Writing CFT correlation functions as AdS scattering
  amplitudes},'' \href{http://dx.doi.org/10.1007/JHEP03(2011)025}{{\em JHEP}
  {\bfseries 03} (2011) 025},
\href{http://arxiv.org/abs/1011.1485}{{\ttfamily arXiv:1011.1485 [hep-th]}}.

\bibitem{Paulos:2011ie}
M.~F. Paulos, ``{Towards Feynman rules for Mellin amplitudes},''
  \href{http://dx.doi.org/10.1007/JHEP10(2011)074}{{\em JHEP} {\bfseries 10}
  (2011) 074},
\href{http://arxiv.org/abs/1107.1504}{{\ttfamily arXiv:1107.1504 [hep-th]}}.

\bibitem{Fitzpatrick:2011ia}
A.~L. Fitzpatrick, J.~Kaplan, J.~Penedones, S.~Raju, and B.~C. van Rees, ``{A
  Natural Language for AdS/CFT Correlators},''
  \href{http://dx.doi.org/10.1007/JHEP11(2011)095}{{\em JHEP} {\bfseries 11}
  (2011) 095},
\href{http://arxiv.org/abs/1107.1499}{{\ttfamily arXiv:1107.1499 [hep-th]}}.

\bibitem{DHoker:1999mqo}
E.~D'Hoker, D.~Z. Freedman, and L.~Rastelli, ``{AdS / CFT four point functions:
  How to succeed at z integrals without really trying},''
  \href{http://dx.doi.org/10.1016/S0550-3213(99)00526-X}{{\em Nucl. Phys.}
  {\bfseries B562} (1999) 395--411},
\href{http://arxiv.org/abs/hep-th/9905049}{{\ttfamily arXiv:hep-th/9905049
  [hep-th]}}.

\bibitem{Galliani:2017jlg}
A.~Galliani, S.~Giusto, and R.~Russo, ``{Holographic 4-point correlators with
  heavy states},'' \href{http://dx.doi.org/10.1007/JHEP10(2017)040}{{\em JHEP}
  {\bfseries 10} (2017) 040},
\href{http://arxiv.org/abs/1705.09250}{{\ttfamily arXiv:1705.09250 [hep-th]}}.

\bibitem{Bombini:2017sge}
A.~Bombini, A.~Galliani, S.~Giusto, E.~Moscato, and R.~Russo, ``{Unitary
  4-point correlators from classical geometries},''
  \href{http://dx.doi.org/10.1140/epjc/s10052-017-5492-3}{{\em Eur. Phys. J.}
  {\bfseries C78} no.~1, (2018) 8},
\href{http://arxiv.org/abs/1710.06820}{{\ttfamily arXiv:1710.06820 [hep-th]}}.

\bibitem{Giusto:2018ovt}
S.~Giusto, R.~Russo, and C.~Wen, ``{Holographic correlators in AdS$_{3}$},''
  \href{http://dx.doi.org/10.1007/JHEP03(2019)096}{{\em JHEP} {\bfseries 03}
  (2019) 096},
\href{http://arxiv.org/abs/1812.06479}{{\ttfamily arXiv:1812.06479 [hep-th]}}.

\bibitem{Bombini:2019vnc}
A.~Bombini and A.~Galliani, ``{AdS$_3$ four-point functions from
  $\frac{1}{8}$-BPS states},''
\href{http://arxiv.org/abs/1904.02656}{{\ttfamily arXiv:1904.02656 [hep-th]}}.

\bibitem{Heydeman:2018dje}
M.~Heydeman, J.~H. Schwarz, C.~Wen, and S.-Q. Zhang, ``{All Tree Amplitudes of
  6D $(2,0)$ Supergravity: Interacting Tensor Multiplets and the $K3$ Moduli
  Space},'' \href{http://dx.doi.org/10.1103/PhysRevLett.122.111604}{{\em Phys.
  Rev. Lett.} {\bfseries 122} no.~11, (2019) 111604},
\href{http://arxiv.org/abs/1812.06111}{{\ttfamily arXiv:1812.06111 [hep-th]}}.

\bibitem{Giusto}
S.~Giusto, R.~Russo, A.~Tyukov, and C.~Wen, ``{Holographic correlators in
  $AdS_3$ without Witten diagrams},'' {\em to appear} .

\bibitem{Dolan:2004mu}
F.~A. Dolan, L.~Gallot, and E.~Sokatchev, ``{On four-point functions of 1/2-BPS
  operators in general dimensions},''
  \href{http://dx.doi.org/10.1088/1126-6708/2004/09/056}{{\em JHEP} {\bfseries
  09} (2004) 056},
\href{http://arxiv.org/abs/hep-th/0405180}{{\ttfamily arXiv:hep-th/0405180
  [hep-th]}}.

\bibitem{Beem:2013sza}
C.~Beem, M.~Lemos, P.~Liendo, W.~Peelaers, L.~Rastelli, and B.~C. van Rees,
  ``{Infinite Chiral Symmetry in Four Dimensions},''
  \href{http://dx.doi.org/10.1007/s00220-014-2272-x}{{\em Commun. Math. Phys.}
  {\bfseries 336} no.~3, (2015) 1359--1433},
\href{http://arxiv.org/abs/1312.5344}{{\ttfamily arXiv:1312.5344 [hep-th]}}.

\bibitem{Baggio:2012rr}
M.~Baggio, J.~de~Boer, and K.~Papadodimas, ``{A non-renormalization theorem for
  chiral primary 3-point functions},''
  \href{http://dx.doi.org/10.1007/JHEP07(2012)137}{{\em JHEP} {\bfseries 07}
  (2012) 137},
\href{http://arxiv.org/abs/1203.1036}{{\ttfamily arXiv:1203.1036 [hep-th]}}.

\bibitem{Deger:1998nm}
S.~Deger, A.~Kaya, E.~Sezgin, and P.~Sundell, ``{Spectrum of D = 6, N=4b
  supergravity on AdS in three-dimensions x S**3},''
  \href{http://dx.doi.org/10.1016/S0550-3213(98)00555-0}{{\em Nucl. Phys.}
  {\bfseries B536} (1998) 110--140},
\href{http://arxiv.org/abs/hep-th/9804166}{{\ttfamily arXiv:hep-th/9804166
  [hep-th]}}.

\bibitem{deBoer:1998kjm}
J.~de~Boer, ``{Six-dimensional supergravity on S**3 x AdS(3) and 2-D conformal
  field theory},'' \href{http://dx.doi.org/10.1016/S0550-3213(99)00160-1}{{\em
  Nucl. Phys.} {\bfseries B548} (1999) 139--166},
\href{http://arxiv.org/abs/hep-th/9806104}{{\ttfamily arXiv:hep-th/9806104
  [hep-th]}}.

\bibitem{deBoer:1998us}
J.~de~Boer, ``{Large N elliptic genus and AdS / CFT correspondence},''
  \href{http://dx.doi.org/10.1088/1126-6708/1999/05/017}{{\em JHEP} {\bfseries
  05} (1999) 017},
\href{http://arxiv.org/abs/hep-th/9812240}{{\ttfamily arXiv:hep-th/9812240
  [hep-th]}}.

\bibitem{Arutyunov:2000by}
G.~Arutyunov, A.~Pankiewicz, and S.~Theisen, ``{Cubic couplings in D = 6 N=4b
  supergravity on AdS(3) x S**3},''
  \href{http://dx.doi.org/10.1103/PhysRevD.63.044024}{{\em Phys. Rev.}
  {\bfseries D63} (2001) 044024},
\href{http://arxiv.org/abs/hep-th/0007061}{{\ttfamily arXiv:hep-th/0007061
  [hep-th]}}.

\bibitem{Zhou:2018sfz}
X.~Zhou, ``{Recursion Relations in Witten Diagrams and Conformal Partial
  Waves},''
\href{http://arxiv.org/abs/1812.01006}{{\ttfamily arXiv:1812.01006 [hep-th]}}.

\bibitem{Eden:2000bk}
B.~Eden, A.~C. Petkou, C.~Schubert, and E.~Sokatchev, ``{Partial
  nonrenormalization of the stress tensor four point function in N=4 SYM and
  AdS / CFT},'' \href{http://dx.doi.org/10.1016/S0550-3213(01)00151-1}{{\em
  Nucl. Phys.} {\bfseries B607} (2001) 191--212},
\href{http://arxiv.org/abs/hep-th/0009106}{{\ttfamily arXiv:hep-th/0009106
  [hep-th]}}.

\bibitem{Lin:2015dsa}
Y.-H. Lin, S.-H. Shao, Y.~Wang, and X.~Yin, ``{Supersymmetry Constraints and
  String Theory on K3},'' \href{http://dx.doi.org/10.1007/JHEP12(2015)142}{{\em
  JHEP} {\bfseries 12} (2015) 142},
\href{http://arxiv.org/abs/1508.07305}{{\ttfamily arXiv:1508.07305 [hep-th]}}.

\bibitem{Arutyunov:2000ima}
G.~Arutyunov and S.~Frolov, ``{On the correspondence between gravity fields and
  CFT operators},'' \href{http://dx.doi.org/10.1088/1126-6708/2000/04/017}{{\em
  JHEP} {\bfseries 04} (2000) 017},
\href{http://arxiv.org/abs/hep-th/0003038}{{\ttfamily arXiv:hep-th/0003038
  [hep-th]}}.

\bibitem{Taylor:2007hs}
M.~Taylor, ``{Matching of correlators in AdS(3) / CFT(2)},''
  \href{http://dx.doi.org/10.1088/1126-6708/2008/06/010}{{\em JHEP} {\bfseries
  06} (2008) 010},
\href{http://arxiv.org/abs/0709.1838}{{\ttfamily arXiv:0709.1838 [hep-th]}}.

\bibitem{Maldacena:2011mk}
J.~Maldacena, ``{Einstein Gravity from Conformal Gravity},''
\href{http://arxiv.org/abs/1105.5632}{{\ttfamily arXiv:1105.5632 [hep-th]}}.

\bibitem{AdS55pt}
V.~Gon\c{c}alves, R.~Pereira, and X.~Zhou, ``{$20'$ Five-Point Function from
  $AdS_5\times S^5$ Supergravity},'' {\em to appear} .

\bibitem{DHoker:1999bve}
E.~D'Hoker, D.~Z. Freedman, S.~D. Mathur, A.~Matusis, and L.~Rastelli,
  ``{Graviton and gauge boson propagators in AdS(d+1)},''
  \href{http://dx.doi.org/10.1016/S0550-3213(99)00524-6}{{\em Nucl. Phys.}
  {\bfseries B562} (1999) 330--352},
\href{http://arxiv.org/abs/hep-th/9902042}{{\ttfamily arXiv:hep-th/9902042
  [hep-th]}}.

\bibitem{Pilch:1984xx}
K.~Pilch and A.~N. Schellekens, ``{Formulae for the Eigenvalues of the
  Laplacian on Tensor Harmonics on Symmetric Coset Spaces},''
\href{http://dx.doi.org/10.1063/1.526101}{{\em J. Math. Phys.} {\bfseries 25}
  (1984) 3455}.

\end{thebibliography}\endgroup
\bibliographystyle{utphys}

\end{document}